\documentclass{article}

\usepackage{amsmath,amssymb}    
\usepackage{color}
\usepackage{graphicx}
\usepackage{subfigure}
\usepackage{cite}               
\usepackage{hyperref}           
\usepackage{multirow,makecell}   

\numberwithin{equation}{section}   

\def \be {\begin{equation}}
\def \ee {\end{equation}}
\def \ba {\begin{array}}
\def \ea {\end{array}}
\def \bea{\begin{eqnarray}}
\def \eea{\end{eqnarray}}

\def \a {\alpha}
\def \b {\beta}

\def \G {\Gamma}
\def \d {\delta}

\def \e {\epsilon}

\def \s {\sigma}

\def \r {\rho}

\def \p {\partial}
\def \f {\frac}

\def \nn {\nonumber}

\def \mc {\mathcal}

\def \lt {\left}
\def \rt {\right}

\def \td {\tilde}

\def \inf {\infty}

\def \Tr {{\textrm{Tr}}}
\def \tr {{\textrm{tr}}}
\def \lag {\langle}
\def \rag {\rangle}

\begin{document}

\title{\textbf{On short interval expansion of R\'enyi entropy}}
\author{
Bin Chen$^{1,2}$\footnote{bchen01@pku.edu.cn}\,
and
Jia-ju Zhang$^{1}$\footnote{jjzhang@pku.edu.cn}
}
\date{}

\maketitle

\begin{center}
{\it
$^{1}$Department of Physics and State Key Laboratory of Nuclear Physics and Technology,\\Peking University, No.~5 Yiheyuan Road, Beijing 100871, P.~R. China\\
\vspace{2mm}
$^{2}$Center for High Energy Physics, Peking University, No.~5 Yiheyuan Road,\\Beijing 100871, P.~R. China
}
\vspace{10mm}
\end{center}

\begin{abstract}
  R\'enyi entanglement entropy provides a new window to study the AdS/CFT correspondence. In this paper we consider the short interval expansion of R\'enyi entanglement entropy in two-dimensional conformal field theory. This amounts to do the operator product expansion of the twist operators. We focus on the vacuum Verma module and consider the quasiprimary operators constructed from the stress tensors. After obtaining the expansion coefficients of the twist operators to level 6 in vacuum Verma module, we compute the leading contributions to the R\'enyi entropy, to order 6 in the short interval expansion. In the case of one short interval on cylinder, we reproduce the first several leading contributions to the R\'enyi entropy. In the case of two short disjoint intervals with a small cross ratio $x$, we obtain  not only  the classical and 1-loop quantum contributions to the R\'enyi entropy to order $x^6$, both of which are in perfect match with the ones found in gravity, but also the leading $1/c$ contributions, which corresponds to 2-loop corrections in the bulk.
\end{abstract}

\baselineskip 18pt
\thispagestyle{empty}

\newpage

\tableofcontents

\section{Introduction}

Entanglement entropy is an important notion of a quantum system, encoding the information of the active degrees of freedom of the system \cite{nielsen2010quantum,petz2008quantum}. It is defined as follows. One can divide the system into two parts, say $A$ and its complement $B$. The Hilbert space of the system is the tensor product of the ones of $A$ and $B$. From the density matrix $\r$ of the whole system  one may obtain the reduced density matrix of $A$ by tracing over the degrees of freedom of $B$
\be
\r_A=\Tr_B \r.
\ee
Then the von Neumann entanglement entropy, or  in short the entanglement entropy, of $A$ and $B$ is defined as
\be
S_A=-\Tr_A \r_A\log \r_A.
\ee
More generally one can define the R\'enyi entanglement entropy, or  in short the R\'enyi entropy, of $A$ and $B$ as
\be
S_A^{(n)}=-\f{1}{n-1} \log \Tr_A \r_A^n.
\ee
It is easy to see that the entanglement entropy and the R\'enyi entropy are related by
\be
S_A=\lim_{n \to 1} S_A^{(n)}.
\ee
Moreover one may choose from the system two subsystems $A$ and $B$ which are not necessarily each other's complement, and define the R\'enyi mutual information of $A$ and $B$
\be
I_{A,B}^{(n)}=S_{A}^{(n)}+S_{B}^{(n)}-S_{A\cup B}^{(n)}.
\ee
In the following, we will write for short $S_n=S_A^{(n)}$ and $I_n=I_{A,B}^{(n)}$ without causing ambiguity.

The entanglement entropy for a conformal field theory (CFT) is of particular interest.  The standard way of computing the entanglement entropy in a quantum field theory is the replica trick \cite{Callan:1994py}. In particular, for a two-dimensional (2D) CFT one could insert the twist operators to impose the nontrivial boundary conditions  in applying the replica trick \cite{Calabrese:2004eu,Calabrese:2009qy}. For a 2D CFT on complex plane the R\'enyi entropy for one interval with length $\ell$ is universal and only depends on the central charge \cite{Calabrese:2004eu}
\be
S_n=\f{c}{6} \lt( 1+\f{1}{n} \rt) \log \f{\ell}{\e},
\ee
with $\e$ being the UV cutoff. But for two and more intervals, the R\'enyi entropy would depend on the details of the CFT \cite{Calabrese:2009ez,Headrick:2010zt,Calabrese:2010he}. For the cases of two disjoint intervals there are some partial results. The analytical results of R\'enyi entropy $S_n, n\geq2$ for free compactified  boson and Ising model  have been presented in \cite{Calabrese:2009ez} and \cite{Calabrese:2010he} respectively. The R\'enyi entropy $S_2$ for a general CFT has been discussed in \cite{Headrick:2010zt}. For a general CFT,  it was proposed in \cite{Headrick:2010zt} that one can use the operator product expansion (OPE) of twist operators to compute the R\'enyi entropy $S_n, n\geq2$. This proposal was generalized to find the leading term of R\'enyi entropy with small cross ratio in \cite{Calabrese:2010he}. Very recently this method was also generalized to higher dimensions in \cite{Cardy:2013nua}.

The direct field theory computation of the entanglement entropy is usually very hard, but for a CFT one may  use the AdS/CFT correspondence \cite{Maldacena:1997re,Gubser:1998bc,Witten:1998qj} to do a simpler  computation in the bulk gravity. For the Einstein gravity, it was firstly proposed by Ryu and Takayanagi \cite{Ryu:2006bv,Ryu:2006ef} that the entanglement entropy could be holographically given by the area of a minimal surface in the bulk, which is homogeneous to $A$. This so-called holographic entanglement entropy has been studied intensely since its proposal, see good reviews \cite{Nishioka:2009un,Takayanagi:2012kg} for complete references. Among various investigations, how to prove the Ryu-Takayanigi (RT) area formula is one of central issues. The early efforts include the works in  \cite{Fursaev:2006ih} and \cite{Casini:2011kv}. The most recent effort from the point of view of generalized gravitational entropy could be found in \cite{Lewkowycz:2013nqa}. One essential point in the proofs is to find the gravitational configurations in applying the replica trick. This turns out to be a subtle issue and has not been well-understood in general cases. However, in AdS$_3$/CFT$_2$ case, the bulk gravitational configurations can be constructed explicitly without trouble. In \cite{Hartman:2013mia,Faulkner:2013yia} the RT formula has been proved in AdS$_3$/CFT$_2$ case.

In the AdS$_3$/CFT$_2$ correspondence, one may consider the large central charge $c$ limit in the CFT side,  corresponding to the weak coupling limit in the gravity side. In this limit, one can separate the R\'enyi entropy into the contributions from the classical, quantum 1-loop, 2-loop, ... parts, with each part being proportional to $c$, $c^0$, $\f{1}{c}$, ..., respectively \cite{Headrick:2010zt}. In the bulk side, the RT area formula gives classical contribution, while the 1-loop contribution comes from the quantum fluctuations around the minimal surface\cite{Barrella:2013wja,Faulkner:2013ana}. For the case with two disjoint intervals, the small cross ratio $x$ expansion of the classical part of R\'enyi entropy to the order $x^6$ has been given in \cite{Hartman:2013mia}. Also using the method in \cite{Faulkner:2013yia} and the 1-loop partition function of graviton in AdS$_3$ \cite{Giombi:2008vd}, the 1-loop correction of R\'enyi entropy to order $x^8$ has been computed in \cite{Barrella:2013wja}. These contributions are expected to be in agreement with the computations in CFT.

In this paper we investigate the short interval expansion of twist operators in more details and compare the results with the bulk ones. According to the AdS/CFT correspondence, the graviton in the bulk corresponds to stress tensor in the boundary theory. For the pure AdS$_3$ gravity, we only need to consider vacuum Verma module. The vacuum Verma module consists of primary identity operator and its decedents which could be constructed by the stress tensors $T(z)$ and $\bar T(\bar z)$. We not only consider the contributions of quasiprimary operators but also the contributions of their derivatives.
Here we discuss two cases. The first one is that there is one interval on cylinder. The other one is of two short disjoint intervals, which may allow small cross ratio expansion. For the first case, we reproduce the universal result to order 6. For the second case we  only consider the contributions to the order $x^6$. We find that the contribution of order $c$ to the R\'enyi entropy matches exactly with the result in \cite{Hartman:2013mia}. Moreover, we read the contributions of order $c^0$ to the R\'enyi entropy, which are in exact agreement with the 1-loop contributions of graviton found in \cite{Barrella:2013wja}. Furthermore, we manage to obtain the contributions of order $\f{1}{c}$, which are expected to match the 2-loop contributions in the bulk.

The remaining of the paper is arranged as follows. In Section~\ref{s2} we introduce the general prescriptions which include a brief review of \cite{Headrick:2010zt,Calabrese:2010he} and our developments. In Section~\ref{s3} we compute the necessary coefficients for future use. In Section~\ref{s4} we present two applications of our method. One is for a single interval on cylinder, and the other is for two intervals on complex plane. In Section~\ref{s5}, we end with conclusion and discussion. We give some useful summation formulas in Appendix~\ref{sa}.

\section{General prescriptions} \label{s2}

In this section we review briefly the short interval expansion proposed in \cite{Headrick:2010zt,Calabrese:2010he}, and moreover present our developments.

Firstly, we present some well known results in a general 2D CFT, which can be found for example in \cite{DiFrancesco:1997nk,Blumenhagen:2009zz}. In a 2D CFT, all the operators could be written in terms of quasiprimary fields and their derivatives. We write the quasiprimary operators as $\phi_i$ with conformal weights $h_i$ and $\bar h_i$. The correlation functions of two and three quasiprimary operators on complex plane $C$ are
\bea
&& \lag \phi_i(z_i,\bar z_i) \phi_j(z_j,\bar z_j)\rag_C=\f{\a_i\d_{ij}}{z_{ij}^{2h_i}\bar z_{ij}^{2\bar h_i}}, \\
&& \lag \phi_i(z_i,\bar z_i) \phi_j(z_j,\bar z_j) \phi_k(z_k,\bar z_k\rag_C=\f{C_{ijk}}
   {z_{ij}^{h_i+h_j-h_k}z_{jk}^{h_j+h_k-h_i}z_{ik}^{h_i+h_k-h_j}\bar z_{ij}^{\bar h_i+\bar h_j-\bar h_k}\bar z_{jk}^{\bar h_j+\bar h_k-\bar h_i}\bar z_{ik}^{\bar h_i+\bar h_k-\bar h_j}} \nn,
\eea
with $z_{ij}\equiv z_i-z_j$ and $\bar z_{ij} \equiv \bar z_i-\bar z_j$. Note that the quasiprimary operators have been orthogonalized but not normalized. From the two-point function there is also
\be
\lag \p^m\bar\p^r \phi_i(z_i,\bar z_i) \p^p\bar\p^q \phi_j(z_j,\bar z_j)\rag_C=
\f{\a_i\d_{ij}(-)^{m+r}(m+p)!(r+q)!C_{2h_i+m+p-1}^{m+p}C_{2\bar h_i+r+q-1}^{r+q}}{z_{ij}^{2h_i+m+p}\bar z_{ij}^{2\bar h_i+r+q}},
\ee
with the binomial coefficient being $C_x^y=\f{\G(x+1)}{\G(y+1)\G(x-y+1)}$. The OPE of two quasiprimary operators could be generally written as
\be \label{e1}
\phi_i(z,\bar z)\phi_j(0,0)
=\sum_k C_{ij}^k \sum_{m,r\geq0} \f{a_{ijk}^m}{m!} \f{\bar a_{ijk}^r}{r!}\f{1}{z^{h_i+h_j-h_k-m} \bar z^{\bar h_i+\bar h_j-\bar h_k-r} } \p^m \bar \p^r \phi_k(0,0),
\ee
where the summation $k$ is over all quasiprimary operators and there are definitions
\be
a_{ijk}^m \equiv \f{C_{h_k+h_i-h_j+m-1}^m}{C_{2h_k+m-1}^m}, ~~~ \bar a_{ijk}^r\equiv\f{C_{\bar h_k+\bar h_i-\bar h_j+r-1}^r}{C_{2\bar h_k+r-1}^r},
~~~ C_{ij}^k\equiv\f{C_{ijk}}{\a_k}.
\ee

We use the replica trick to calculate the R\'enyi entanglement entropy. The trick requires us to make $n$ copies of the original CFT, which we call $CFT_n$. When there are $N$ disjoint intervals $A$, we denote the Riemann surface where $CFT_n$ resides as $\mc R_{n,N}$. The properties of $CFT_1$ is just what we have reviewed above, and for the general $CFT_n$ there are similar properties. The $CFT_n$ has central charge $nc$ with $c$ being the central charge of $CFT_1$, and the stress tensors are
\be
\sum_{j=0}^{n-1} T(z_j), ~~~ \sum_{j=0}^{n-1} \bar T(\bar z_j)
\ee
where $T(z_j)$, $\bar T(\bar z_j)$ are the stress tensors of the $j$-th copy the original CFT and $z_j$ is the coordinate of the $j$-th copy of the Riemann surface $\mc R_{n,N}$. We denote the linear independent quasiprimary operators of $CFT_n$ as $\Phi_K(z,\bar z)$ with conformal wights $h_K$ and $\bar h_K$. The product of quasiprimary operators in each copy forms a quasiprimary operator of $CFT_n$,
\be
\Phi_K(z,\bar z)=\prod_{j=0}^{n-1} \phi_{k_j}(z_j,\bar z_j),
\ee
and in this case there are
\be \label{e33}
K=\{k_j\}, ~~~ \a_K=\prod_{j=0}^{n-1}\a_{k_j}, ~~~ h_K=\sum_{j=0}^{n-1} h_{k_j}, ~~~ \bar h_K=\sum_{j=0}^{n-1} \bar h_{k_j}.
\ee
Note that not all of the quasiprimary operators of $CFT_n$ could be written in the above form \cite{Headrick:2010zt}, and we will see examples (\ref{eee50}) in the following sections.

To compute the $n$-th R\'enyi entanglement entropy of the original CFT, we have to compute the partition function of $CFT_n$ on $\mc R_{n,N}$. There are two different views for the computation \cite{Calabrese:2004eu}. The first view is to compute it directly on $\mc R_{n,N}$. This requires us to consider the nontrivial boundary conditions of $CFT_n$ on $\mc R_{n,N}$. Then we have
\be
\Tr \r_A^n=\f{Z_n(A)}{Z^n},
\ee
 where $Z=Z_1(A)$. The second view is to replace the boundary conditions with the insertions of the twist operators $\s(z,\bar z)$, $\td \s(z,\bar z)$ at the boundary of each interval, and at the same time replace the Riemann surface $\mc R_{n,N}$ with the complex plane $C$. From the second viewpoint, each copy of the CFT relates to each other only through the twist operators. Both of the twist operators have conformal weights \cite{Calabrese:2004eu}
\be h=\bar h=\f{c}{24} \lt( n-\f{1}{n} \rt).\ee
If we denote $A=[z_1,z_2]\cup\cdots\cup[z_{2N-1},z_{2N}]$, we have
\be
\Tr \r_A^n=
\lag \s(z_{2N},\bar z_{2N})\td \s(z_{2N-1},\bar z_{2N-1}) \cdots \s(z_{2},\bar z_{2})\td \s(z_{1},\bar z_{1}) \rag_{C}.
\ee
For example, when $N=1$ and $A=[0,\ell]$, we have
\be
\Tr \r_A^n=\lag \s(\ell,\ell)\td \s(0,0) \rag_{C}=c_n \ell^{-\f{c}{6}\lt( n-\f{1}{n} \rt)},
\ee
from which the R\'enyi entropy for one interval could be read\cite{Calabrese:2004eu}
\be
S_n=\f{c}{6}\lt( 1+\f{1}{n} \rt)\log\f{\ell}{\e},
\ee
with $\e$ being the UV cutoff.

When the intervals are short, similar to (\ref{e1}), we have the OPE of the twist operators in $CFT_n$
\be \label{e2}
\s(z,\bar z)\td \s(0,0)
=c_n \sum_K d_K \sum_{m,r\geq0} \f{a_K^m}{m!}\f{\bar a_K^r}{r!}\f{1}{z^{2h-h_K-m}\bar z^{2\bar h-\bar h_K-r}} \p^m \bar \p^r \Phi_K(0,0),
\ee
with the summation $K$ being over all the independent quasiprimary operators of $CFT_n$. In (\ref{e2})  there are  definitions
\bea
&& a_K^m\equiv \f{C_{h_K+m-1}^m}{C_{2h_K+m-1}^m}, ~~~ \bar a_K^r\equiv\f{C_{\bar h_K+r-1}^r}{C_{2\bar h_K+r-1}^r}.
\eea
We denote $z=\ell$ and keep in mind that $d_K$ is independent of $\ell$. For a quasiprimary operator $\Phi_K$, the OPE coefficient is
\be \label{ck}
C_K=c_n \ell^{-\f{c}{6}\lt( n-\f{1}{n} \rt)}d_K,
\ee
and the OPE coefficient of  its derivatives $\p^m \bar \p^r \Phi_K$ is
\be \label{ckmr}
C_K^{(m,r)}=c_n \ell^{-\f{c}{6}\lt( n-\f{1}{n} \rt)+m+r}  d_K \f{a_K^m}{m!}\f{\bar a_K^r}{r!}.
\ee

\begin{figure}
\centering
\subfigure[]{\includegraphics[width=6.2cm]{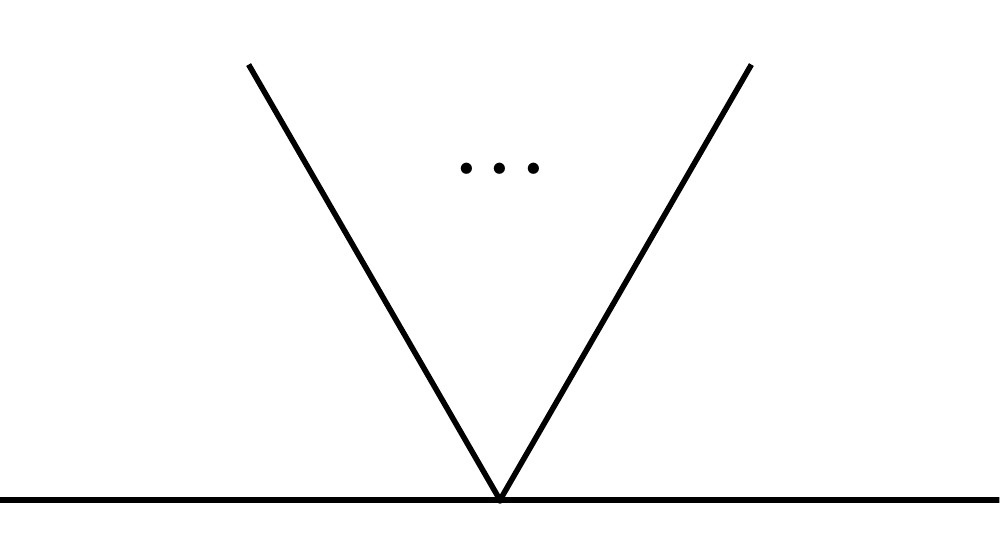} \label{oo}}
\subfigure[]{\includegraphics[width=6.2cm]{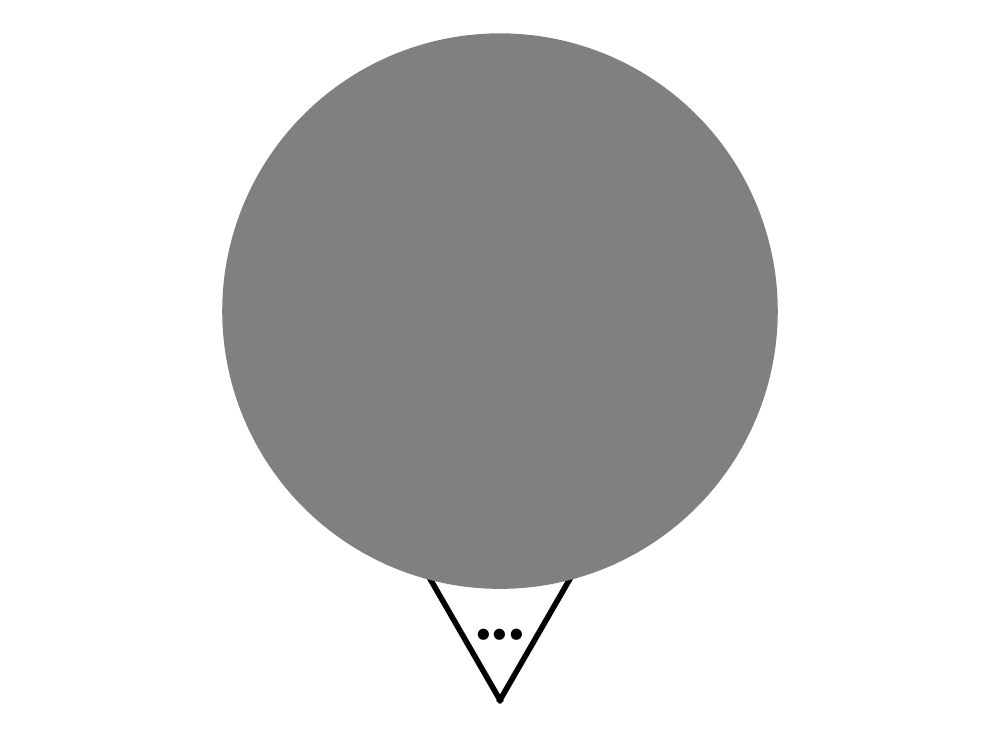} \label{pp}}
\caption{(a) OPE vertex of twist operators; (b) Expectation value of an operator on $\mc R_{n,1}$.} \label{f1}
\end{figure}

Note that the OPE of the twist operators could be represented by a  diagram Fig.~\ref{oo}. In Fig.~\ref{oo} the left and right sides of the horizontal line represent the twist operators $\s$ and $\td \s$, and each of the other lines represents a nonidentity operator in  different copies of the CFT. The cross point of these lines is the coupling of the vertex and represents the OPE coefficient (\ref{ck}) or (\ref{ckmr}) of the twist operators. When there is only the horizontal line it represents the identity operator $\Phi_K=1$, and we have $d_1=1$ according to our normalization. Note that we have not assigned the operator that each line represents, and each diagram may represent many different processes.

When there is one interval $A=[0,\ell]$, we consider the expectation value of one quasiprimary operator $\Phi_K(z,\bar z)$ on $\mc R_{n,1}$, and then we have \cite{Calabrese:2010he}
\be \label{e29}
\f{Z_n(A)}{Z^n}\lag \Phi_K(z,\bar z) \rag_{\mc R_{n,1}}=\lag \Phi_K(z,\bar z) \s(\ell,\ell)\td \s(0,0) \rag_{C}.
\ee
Then using (\ref{e2}) and the orthogonality of quasiprimary operators of $CFT_n$ we have
\be \label{e3}
d_K=\f{1}{\a_K\ell^{h_K+\bar h_K}} \lim_{z\to\inf}z^{2 h_K}\bar z^{2\bar h_K}\lag \Phi_K(z,\bar z) \rag_{\mc R_{n,1}},
\ee
with $\a_K$ being a normalization coefficient.

The computation of the expectation value of an operator on $\mc R_{n,1}$ could be described schematically by Fig~\ref{pp}. Each line represents a nonidentity operator in  different copies of the CFT and the shaded region represents the contraction of these operators, and the expectation value of the operator is the summation over all possible contractions. The diagrams will be explained further in Section~\ref{s3} with concrete examples.

Therefore, the key ingredients in the OPE of twist operators is to calculate the coefficients $\a_K$ and $d_K$. For a general CFT and its Verma modules, it could be tedious to determine these coefficients. For the vacuum Verma module, it is tractable, at least for the first few levels.

\section{The coefficients $\a_K$ and $d_K$} \label{s3}

In this section we consider the coefficients $\a_K$ and $d_K$ of the first several quasiprimary operators in vacuum Verma module.

For the $CFT_1$, to  level 6 the partition function of the holomorphic part of the vacuum Verma module is
\be
\tr x^{L_0}=\prod_{m=2}^\inf\f{1}{1-x^m}=1+x^2+x^3+2x^4+2x^5+4x^6+O(x^7).
\ee
So the number of linear independent holomorphic quasiprimary operators at each level $L_0$ is
\begin{center}
\begin{tabular}{c|c|c|c|c|c|c|c|c}
  $L_0$ & 0 & 1 & 2 & 3 & 4 & 5 & 6 & $\cdots$  \\\hline
 \# & 1 & 0 & 1 & 0 & 1 & 0 & 2 & $\cdots$  \\
\end{tabular}
\end{center}
Explicitly these holomorphic quasiprimary operators are listed as  follows.
\begin{itemize}
  \item At level 0, it is the identity operator 1.
  \item At level 2, there is one quasiprimary operator the stress tensor $T$.
  \item At level 4, it is $\mc O=(TT)-\f{3}{10}\p^2T$.
  \item At level 6, they are $\mc Q=(\p T\p T)-\f{2}{9}\p^2(TT)+\f{1}{42}\p^4 T$ and $\mc R=\mc P+\f{9(14c+43)}{2(70c+29)}\mc Q$, with $\mc P=(T(TT))-\f{1}{4}\p^2(TT)+\f{1}{56}\p^4 T$.
\end{itemize}
We use the notation $(A B)(z)$ representing the normal ordering of two operators $A(z)$ and $B(z)$. Note that at level 6, $\mc P(z)$ and $\mc Q(z)$ are not orthogonal. After using the Gram-Schmidt orthogonalization process, we get the orthogonalized operators $\mc Q(z)$ and $\mc R(z)$.

The normalization factor $\a_k$ of an Hermite operator $\phi_k(z,\bar z)$ can be computed easily. Firstly one define the state
\be
|k\rag \equiv \phi_k(0,0) |0\rag,
\ee
with $|0\rag$ being the vacuum state of the CFT on $C$, and then
\be
\a_k=\lag k| k\rag.
\ee
For example, for the operator $\mc O(z)$ we have
\be
|\mc O\rag= \lt( L_{-2}L_{-2}-\f{3}{5}L_{-4} \rt) |0\rag,
\ee
and then
\be
\a_{\mc O}=\f{c(5c+22)}{10}.
\ee
Similarly, for other quasiprimary operators, their normalization factors  are respectively
\be
\a_1=1, ~~~ \a_T=\f{c}{2}, ~~~ \a_{\mc Q}=\f{4c(70c+29)}{63}, ~~~ \a_{\mc R}=\f{3c(2c-1)(5c+22)(7c+68)}{4(70c+29)}.
\ee

There are also the antiholomorphic quasiprimary operators $\bar T$, $\bar {\mc O}$, $\bar {\mc Q}$ and $\bar {\mc R}$, as well as the quasiprimary operators with mixing holomorphic and antiholomorphic parts. We consider the partition function of vacuum Verma module
\be
\tr' x^{L_0+\bar L_0}=(1+x^2+x^4+2x^6+O(x^8))^2=1+2x^2+3x^4+6x^6+O(x^8),
\ee
and here $\tr'$ counts the quasiprimary operators. So the number of quasiprimary operators at each level $L_0+\bar L_0$ is
\begin{center}
\begin{tabular}{c|c|c|c|c|c|c|c|c}
  $L_0+\bar L_0$ & 0 & 1 & 2 & 3 & 4 & 5 & 6 & $\cdots$  \\\hline
 \# &1 &0 & 2 & 0 & 3 & 0 & 6 & $\cdots$  \\
\end{tabular}
\end{center}
\begin{itemize}
  \item At level 0, it is 1.
  \item At level 2, they are $T$ and $\bar T$.
  \item At level 4, they are $\mc O$, $\bar {\mc O}$ and $T\bar T$.
  \item At level 6, they are $\mc Q$, $\mc R$, $\bar{\mc Q}$, $\bar{\mc R}$, $T\bar{\mc O}$ and $\bar T{\mc O}$.
\end{itemize}
Note that here the quasiprimary operators are just trivial multiplications of the holomorphic and antiholomorphic parts,  because that the OPE of $T$ and $\bar T$ has no singular terms. This observation will simplify our calculations in Section~\ref{s4}.

For $CFT_n$, the holomorphic partition function of the vacuum Verma module is
\be
\tr x^{L_0}=\prod_{m=2}^\inf\f{1}{(1-x^m)^n}=1+n x^2+n x^3+\f{n(n+3)}{2}x^4+n(n+1)x^5+\f{n(n+1)(n+11)}{6}x^6+O(x^7).
\ee
The number of holomorphic quasiprimary operators at each level $L_0$ is
\begin{center}
\begin{tabular}{c|c|c|c|c|c|c|c|c}
  $L_0$ & 0 & 1 & 2 & 3 & 4 & 5 & 6 & $\cdots$ \\ \hline
  \# & 1 & 0 & $n$ & 0 & $\f{n(n+1)}{2}$ & $\f{n(n-1)}{2}$ & $\f{n(n+1)(n+5)}{6}$ & $\cdots$ \\
\end{tabular}
\end{center}
The quasiprimary operators are listed as below.
\begin{center} \begin{tabular}{c|c|c|c}
       $L_0 $ & quasiprimary operators & degeneracies & \# \\\hline
       0 & 1 & 1 & 1\\ \hline
       2 & $T(z_j)$ & $n$ & $n$ \\\hline
       \multirow{2}*{4} & $T(z_{j_1})T(z_{j_2})$ with $j_1 < j_2$ & $\f{n(n-1)}{2}$ & \multirow{2}*{$\f{n(n+1)}{2}$} \\ \cline{2-3}
         & $\mc O(z_j)$ & $n$ & \\\hline
       5 & $\mc S_{j_1 j_2}(z)$ with $j_1 < j_2$ & $\f{n(n-1)}{2}$ & $\f{n(n-1)}{2}$ \\ \hline
         & $T(z_{j_1})T(z_{j_2})T(z_{j_3})$ with $j_1 < j_2<j_3$ & $\f{n(n-1)(n-2)}{6}$ & \\ \cline{2-3}
         & $T(z_{j_1}) \mc O(z_{j_2})$ with $j_1 \neq j_2$ & $n(n-1)$ & \\ \cline{2-3}
       6 & $\mc U_{j_1j_2}(z)$ with $j_1 < j_2$ & $\f{n(n-1)}{2}$ &  $\f{n(n+1)(n+5)}{6}$   \\ \cline{2-3}
         & $\mc Q(z_j)$ & $n$ & \\ \cline{2-3}
         & $\mc R(z_j)$ & $n$ & \\ \hline
       $\cdots$ & $\cdots$ & $\cdots$ & $\cdots$
\end{tabular} \end{center}
Note that the $j$'s listed above vary as $0 \leq j \leq n-1$, and also there are
\bea \label{eee50}
&& \mc S_{j_1 j_2}(z)=T(z_{j_1})i\p T(z_{j_2})-i\p T(z_{j_1})T(z_{j_2}),  \nn\\
&& \mc U_{j_1j_2}(z)=\f{5}{9} \p T(z_{j_1})\p T(z_{j_2})- \f{2}{9} \p^2 T(z_{j_1})T(z_{j_2})-\f{2}{9} T(z_{j_1})\p^2 T(z_{j_2}).
\eea
As we have stated in Section~\ref{s2}, they are examples of quasiprimary operators that could not be written as multiplications of quasiprimary operators at each copy. The coefficient $\a_K$ for these operators could be calculated easily
\bea
&& \a_{TT}=\f{c^2}{4}, ~~~ \a_{\mc S}=2c^2, ~~~ \a_{TTT}=\f{c^3}{8},  \nn\\
&& \a_{T\mc O}=\f{c^2(5c+22)}{20}, ~~~ \a_{\mc U}=\f{20c^2}{9}.
\eea
For example, when we calculate $\a_{\mc S}$, we use
\be
| \mc S_{j_1j_2}\rag=i \lt( L^{(j_1)}_{-2}L^{(j_2)}_{-3}-L^{(j_1)}_{-3}L^{(j_2)}_{-2} \rt) |0\rag.
\ee

To compute $d_K$ we consider the multivalued transformation \cite{Calabrese:2004eu,Calabrese:2010he}
\be
z \to f(z)=\lt( \f{z-\ell}{z} \rt)^{1/n},
\ee
which maps the Riemann surface $\mc R_{n,1}$ with the coordinate $z$ to the complex plane $C$ with the coordinate $f$. With some efforts, we can get $d_K$ for various operators listed above,
\bea \label{e16}
&& d_1=1, ~~~
   d_T=\f{n^2-1}{12n^2}, ~~~
   d_{TT}^{j_1j_2}=\f{1}{8n^4c}\f{1}{s^4_{j_1j_2}}+\f{(n^2-1)^2}{144n^4},  \nn\\
&& d_{\mc O}=\f{(n^2-1)^2}{288n^4}, ~~~
   d_{\mc S}^{j_1j_2}=\f{1}{16n^5 c}\f{c_{j_1j_2}}{s^5_{j_1j_2}}, \nn\\
&& d_{TTT}^{j_1j_2j_3}=-\f{1}{8n^6c^2}\f{1}{s^2_{j_1j_2}s^2_{j_2j_3}s^2_{j_1j_3}}
                       +\f{n^2-1}{96n^6c} \lt( \f{1}{s^4_{j_1j_2}}+\f{1}{s^4_{j_2j_3}}+\f{1}{s^4_{j_1j_3}} \rt)
                       +\f{(n^2-1)^3}{1728n^6}, \nn\\
&& d_{T\mc O}^{j_1j_2}=\f{n^2-1}{96n^6c}\f{1}{s^4_{j_1j_2}}+\f{(n^2-1)^3}{3456n^6}, ~~~
   d_{\mc Q}=-\f{(n^2-1)^2\lt( 2(35c+61)n^2-93 \rt)}{5760n^6(70c+29)}, \nn\\
&& d_{\mc R}=\f{(n^2-1)^3}{10368n^6}, ~~~
   d_{\mc U}^{j_1j_2}=\f{9}{128n^6c}\f{1}{s^6_{j_1j_2}}-\f{n^2+9}{160n^6c}\f{1}{s^4_{j_1j_2}}-\f{(n^2-1)^2}{2880n^4}.
\eea
Here we have defined $s_{j_1j_2}\equiv\sin\f{\pi(j_1-j_2)}{n}$ and $c_{j_1j_2}\equiv\cos\f{\pi(j_1-j_2)}{n}$ for simplicity. The coefficients for the antiholomorphic part and the mixing of both holomorphic and antiholomorphic parts could be got easily from the above results.

Before we  give the derivations of some  coefficients above, we firstly define some shorthands to make our equations simpler. We denote
\bea
&& f=f(z), ~~~ f'=f'(z), ~~~ f''=f''(z), ~~~ \cdots  \nn\\
&& f_{j}=f(z_j), ~~~ f'_j=f'(z_j), ~~~ f''_j=f''(z_j), ~~~\cdots.
\eea
Note that $f$ is multivalued, when $z \to \inf$ we have $f_j \to e^{\f{2\pi i}{n}j}$, $f'_j \to \f{\ell e^{\f{2\pi i}{n}j}}{nz^2}$, $\cdots$. We also denote the Schwarz derivative as
\be
s(z) \equiv \{ f(z);z \} = \f{f'''(z)}{f'(z)}-\f{3}{2} \lt( \f{f''(z)}{f'(z)} \rt)^2,
\ee
and the shorthands
\bea
&& s=s(z), ~~~ s'=s'(z), ~~~ s''=s''(z), ~~~ \cdots  \nn\\
&& s_{j}=s(z_j), ~~~ s'_j=s'(z_j), ~~~ s''_j=s''(z_j), ~~~\cdots.
\eea
Note that $s(z)$ is not multivalued, and so the subscript $j$ could be omitted.

We know that the stress tensor $T$ transforms as
\be \label{e14}
T(z)=f'^2T(f)+\f{c}{12}s,
\ee
and so we have
\be
\lag T(z_j) \rag_{\mc R_{n,1}}= \lag f_j'^2T(f_j)+\f{c}{12}s \rag_{C}=\f{c}{12}s=\f{(n^2-1)c\ell^2}{24n^2z^2(z-\ell)^2},
\ee
from which we get $d_T$. From $T(z)T(w)$ OPE, we can get the transformation of $\mc O$ as
\be \label{e15}
\mc O(z)=f'^4\mc O(f)+\f{5c+22}{30}s \lt( f'^2 T(f)+\f{c}{24}s \rt),
\ee
from which we get $d_{\mc O}$. The coefficients of $d_{\mc Q}$ and $d_{\mc R}$ could be computed in similar ways. The transformation of $\mc Q$ is nontrivial, and we will not give its explicit form. But the transformation for $\mc R$ is relatively simple,
\be
\mc R(z)=f'^6\mc R(f)+\f{(2c-1)(7c+68)}{70c+29}s \lt( \f{5}{4} f'^4\mc O(f)+\f{5c+22}{48}s \lt( f'^2T(f)+\f{c}{36}s \rt) \rt).
\ee
When we calculate $d_{TT}^{j_1j_2}$ and $d_{T\mc O}^{j_1j_2}$, we have to firstly use (\ref{e14}), (\ref{e15}) and get respectively
\bea
&& \lag T(z_{j_1})T(z_{j_2}) \rag_{\mc R_{n,1}}=f_{j_1}'^2f_{j_2}'^2 \lag T(f_{j_1})T(f_{j_2})\rag_{C}+\f{c^2}{144}s^2, \nn\\
&& \lag T(z_{j_1})\mc O(z_{j_2}) \rag_{\mc R_{n,1}}=\f{5c+22}{30}
        \lt(  f_{j_1}'^2f_{j_2}'^2 s \lag T(f_{j_1})T(f_{j_2})\rag_{C}+\f{c^2}{288}s^3  \rt).
\eea
To calculate $d_{\mc S}^{j_1j_2}$ and $d_{\mc U}^{j_1j_2}$, we have to use the formulas
\bea
&& \p T(z)=f'^3\p T(f)+2f'f''T(f)+\f{c}{12}s',  \nn\\
&& \p^2T(z)=f'^4\p^2T(f)+5f'^2f''\p T(f)+2(f''^2+f'f''')T(f)+\f{c}{12}s''.
\eea
At last, to get $d_{TTT}^{j_1j_2j_3}$ we have to firstly get
\bea
&& \lag T(z_{j_1}) T(z_{j_2}) T(z_{j_3})\rag_{\mc R_{n,1}}=
f_{j_1}'^2 f_{j_2}'^2 f_{j_3}'^2 \lag T(f_{j_1})T(f_{j_2})T(f_{j_3}) \rag_{C}  \nn\\
&& \phantom{ \lag T(z_{j_1}) T(z_{j_2}) T(z_{j_3})\rag_{\mc R_{n,1}}=}
+\f{c}{12}s\lt(  f_{j_1}'^2f_{j_2}'^2 \lag T(f_{j_1})T(f_{j_2})\rag_{C}+f_{j_2}'^2f_{j_3}'^2 \lag T(f_{j_2})T(f_{j_3})\rag_{C}+f_{j_1}'^2f_{j_3}'^2 \lag T(f_{j_1})T(f_{j_3})\rag_{C} \rt) \nn\\
&& \phantom{ \lag T(z_{j_1}) T(z_{j_2}) T(z_{j_3})\rag_{\mc R_{n,1}}=}
+\f{c^3}{1728}s^3.
\eea

\begin{figure}
\centering
\subfigure[]{\includegraphics[width=6.2cm]{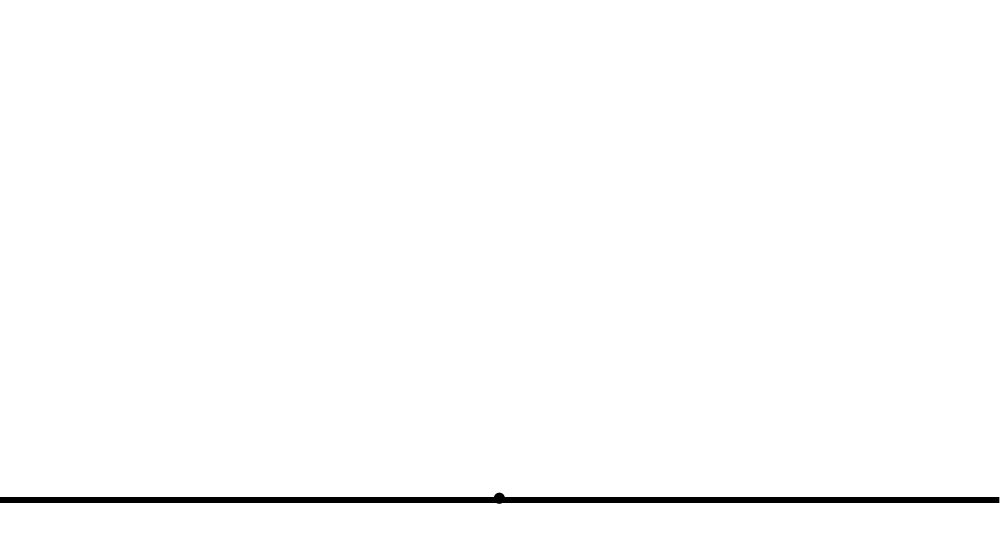} \label{o0}}
\subfigure[]{\includegraphics[width=6.2cm]{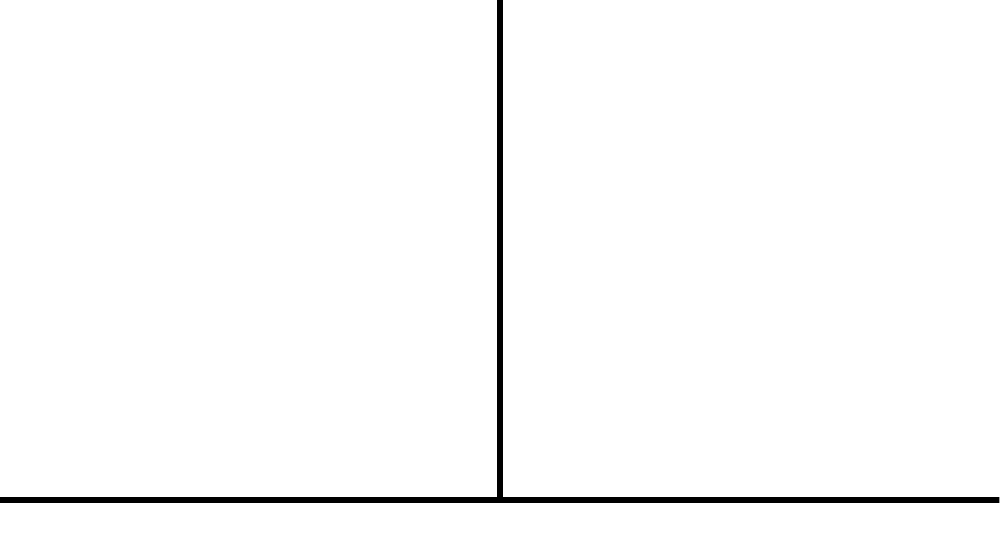} \label{o1}}
\subfigure[]{\includegraphics[width=6.2cm]{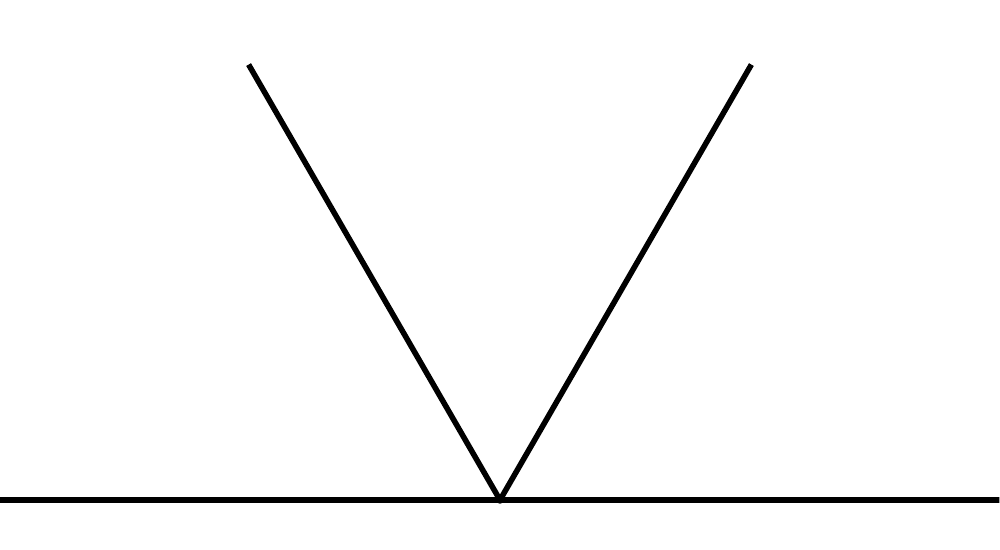} \label{o2}}
\subfigure[]{\includegraphics[width=6.2cm]{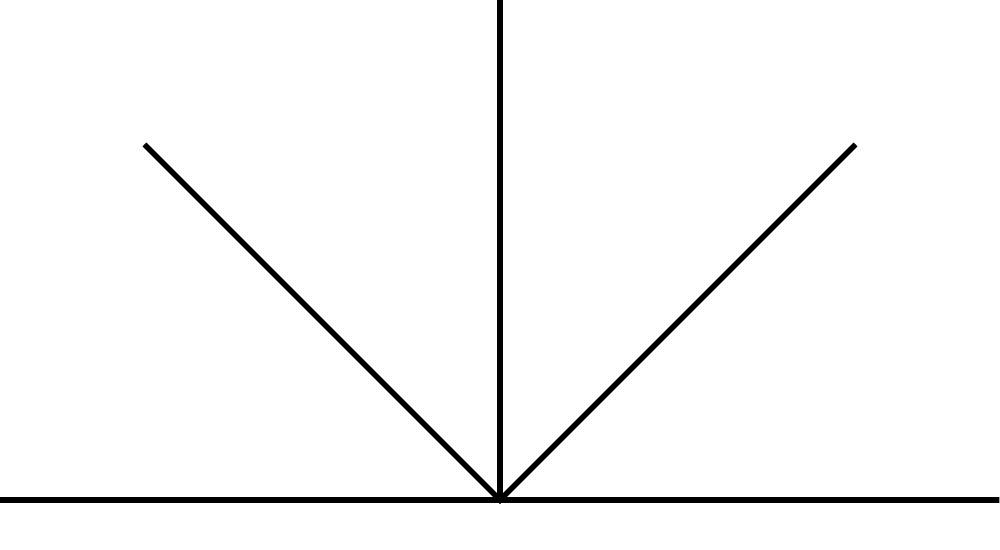} \label{o3}}
\caption{The diagrams for OPE of twist operators} \label{ope}
\end{figure}

\begin{figure}
\centering
\subfigure[]{\includegraphics[width=6.2cm]{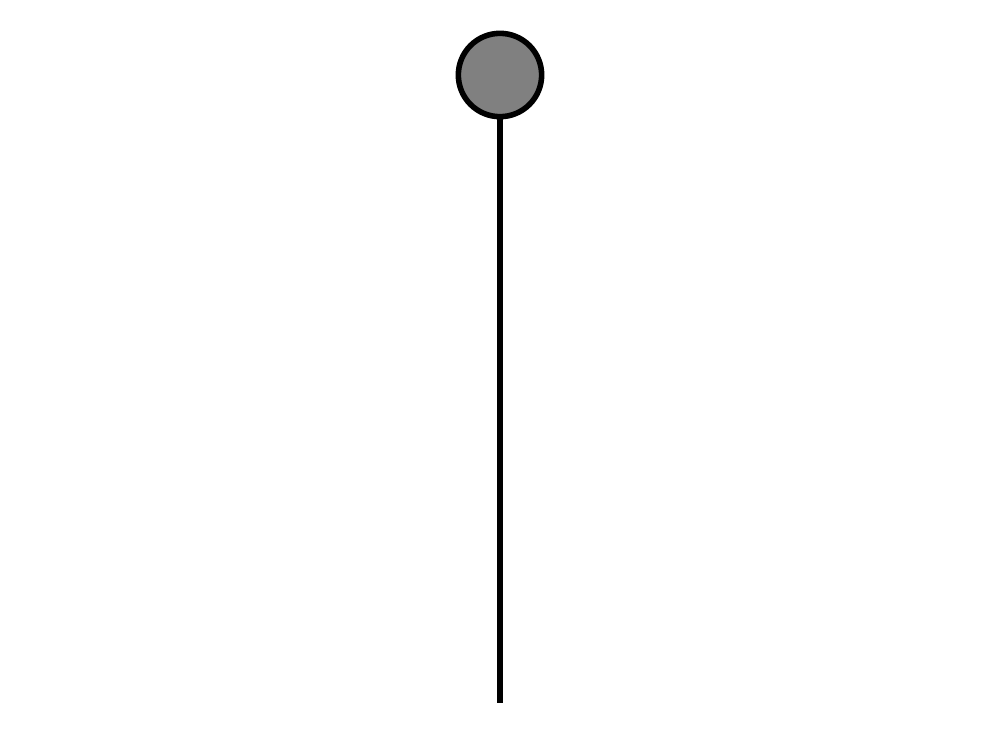} \label{p1}}
\subfigure[]{\includegraphics[width=6.2cm]{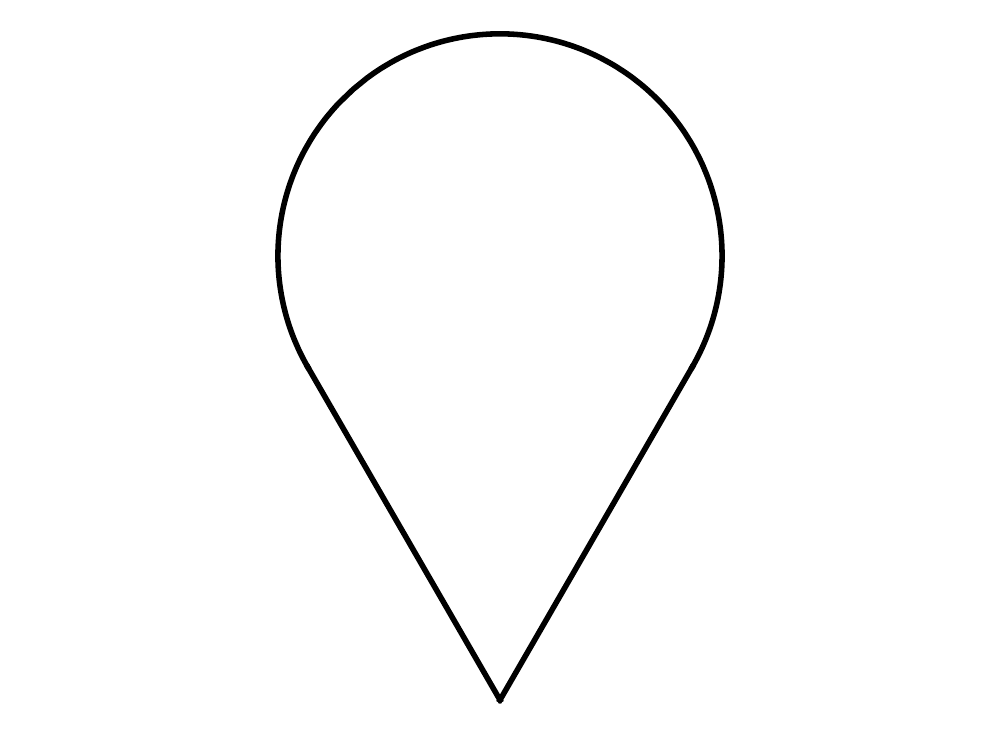} \label{p4}}
\subfigure[]{\includegraphics[width=6.2cm]{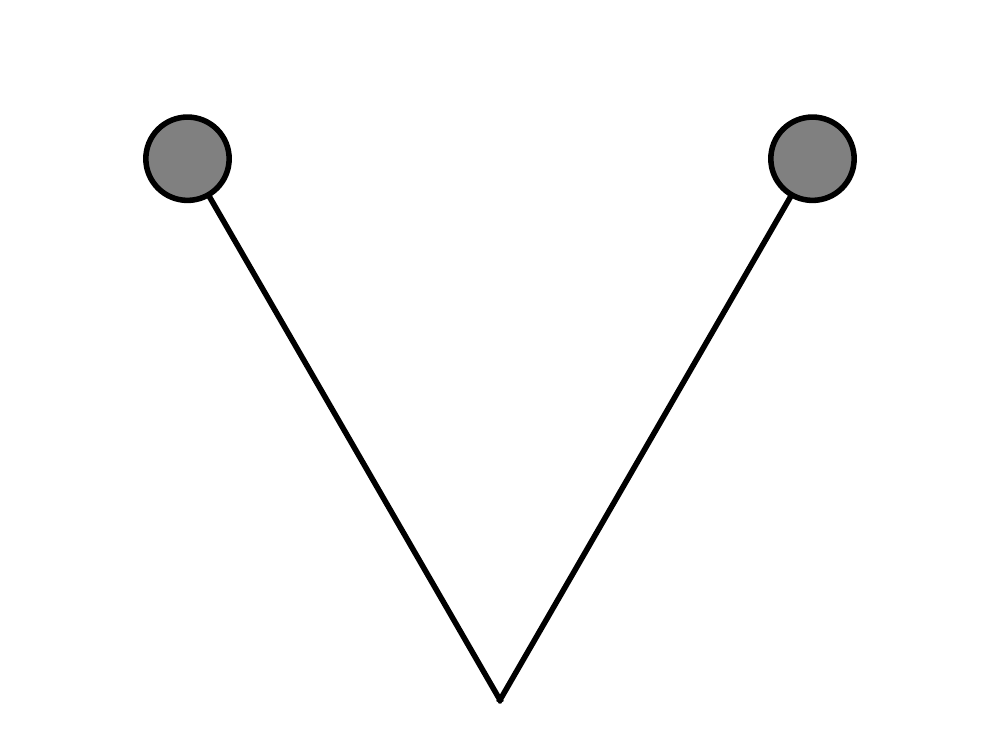} \label{p2}}
\subfigure[]{\includegraphics[width=6.2cm]{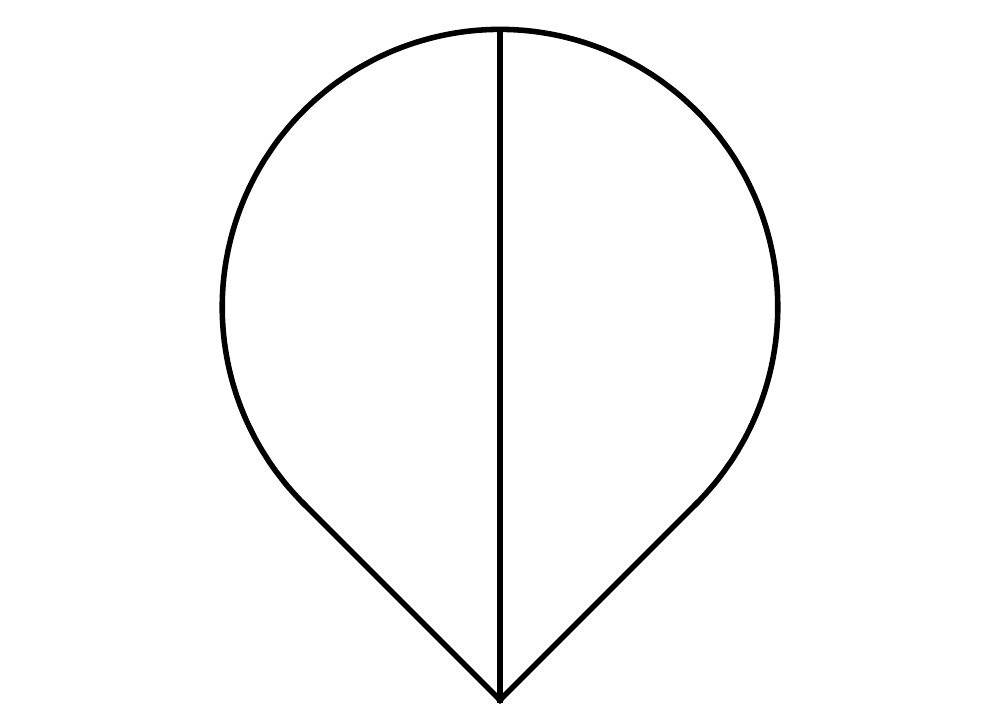} \label{p5}}
\subfigure[]{\includegraphics[width=6.2cm]{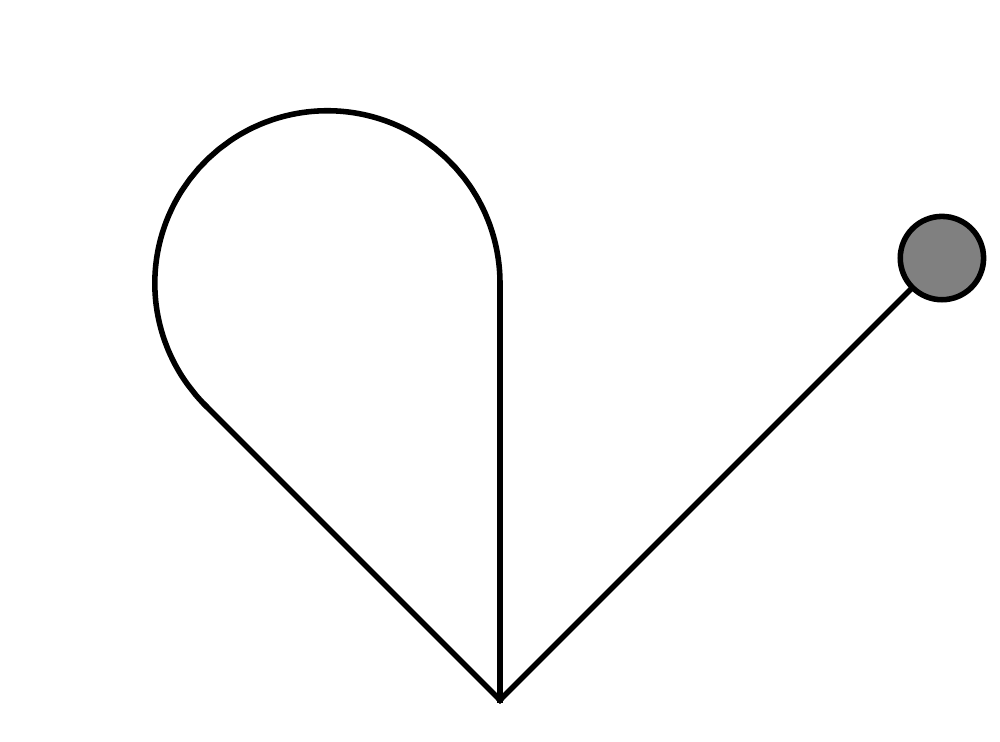} \label{p6}}
\subfigure[]{\includegraphics[width=6.2cm]{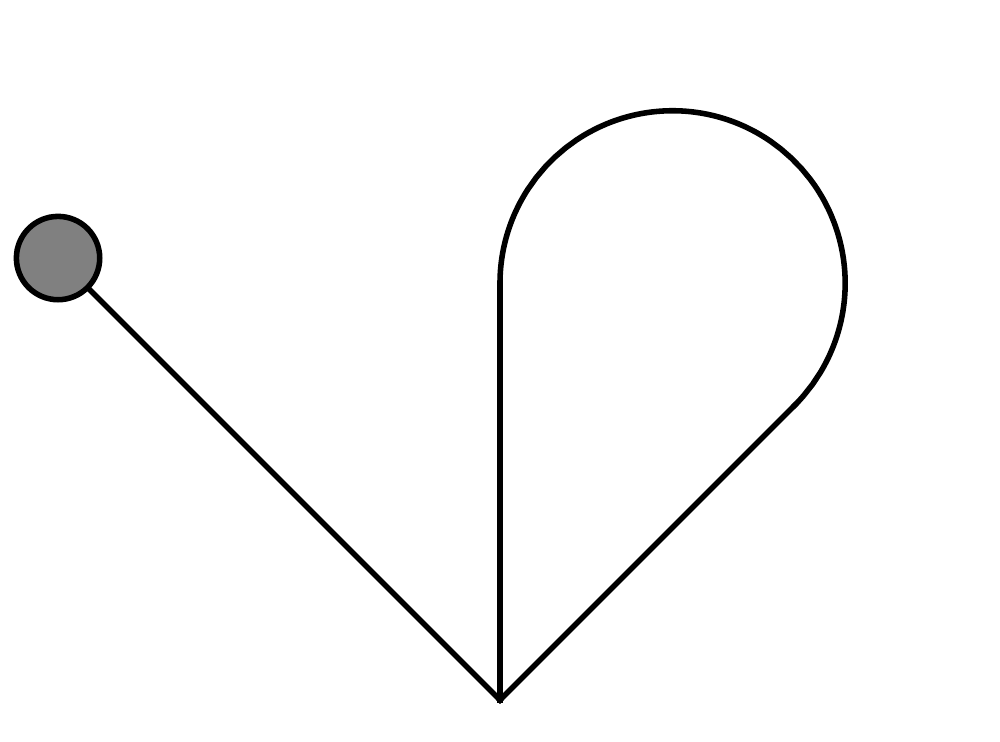} \label{p7}}
\subfigure[]{\includegraphics[width=6.2cm]{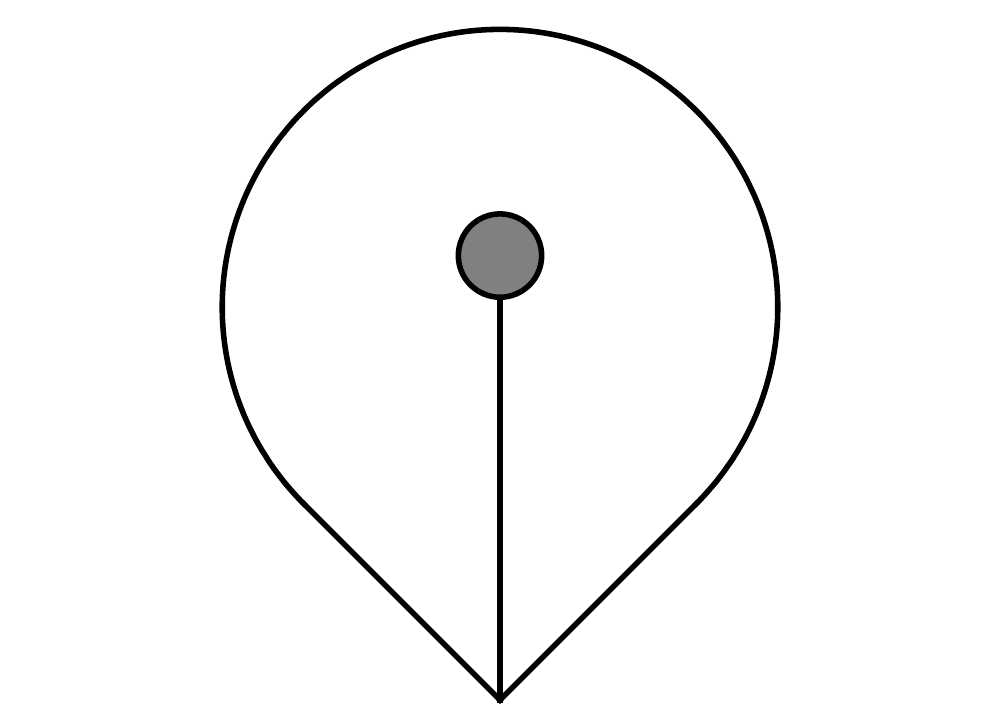} \label{p8}}
\subfigure[]{\includegraphics[width=6.2cm]{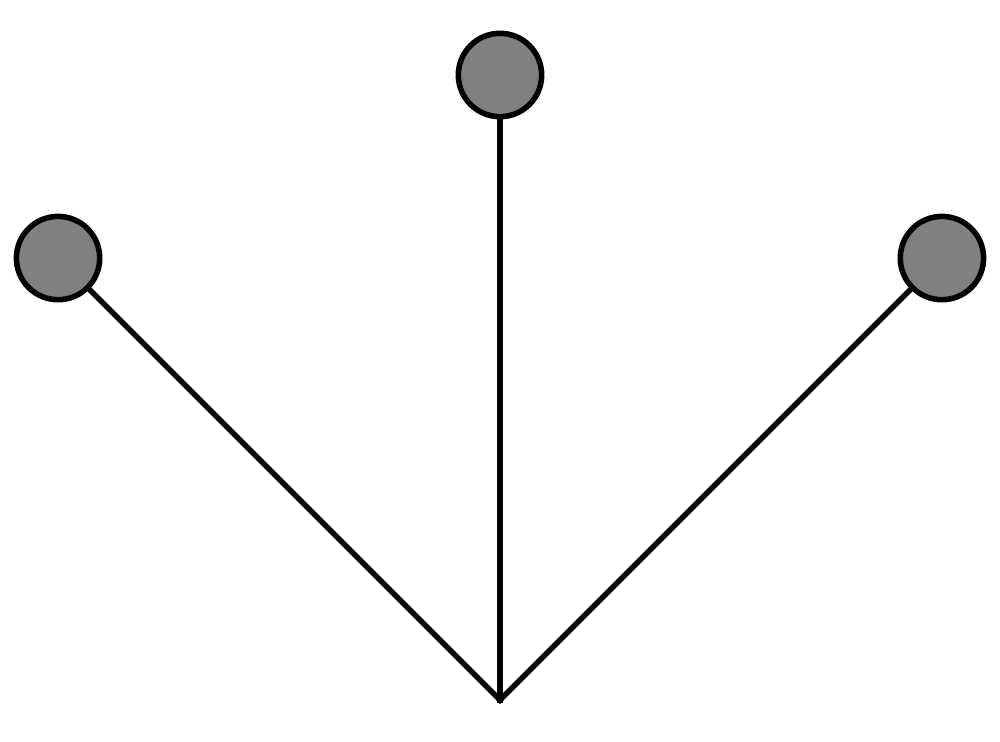} \label{p3}}
\caption{The diagrams for the expectation values of the operators on $\mc R_{n,1}$} \label{tu}
\end{figure}

As we have said, (\ref{e3}) amounts to calculate the  diagram \ref{oo} through the diagram \ref{pp}. More explicitly, to level 6 of vacuum Verma module, the diagrams that appear in the previous calculations  are listed in Fig.~\ref{ope} and Fig.~\ref{tu}.
\begin{itemize}
  \item The computation of the coefficient $d_1$ needs Fig.~\ref{o0} and is trivial.
  \item The computations of the coefficients $d_T$, $d_{\mc O}$, $d_{\mc Q}$ and $d_{\mc R}$ need Fig.~\ref{o1}, or more precisely the diagram Fig.~\ref{p1}. In these cases, there is only one quasiprimary operator from one copy of the CFT so that the small gray disk in Fig.~\ref{p1} means the expectation value of the single operator on $\mc R_{n,1}$.
  \item The computations of the coefficients $d_{TT}^{j_1 j_2}$ and $d_{T\mc O}^{j_1 j_2}$ correspond to Fig.~\ref{o2}, or more precisely the diagrams Fig.~\ref{p4} and \ref{p2}. In both cases there are two operators from two different copies of the CFT. Note that there are two terms in both $d_{TT}^{j_1 j_2}$ and $d_{T\mc O}^{j_1 j_2}$, receiving contributions from the diagrams Fig.~\ref{p4} and \ref{p2} respectively.
  \item The computations of the coefficients $d_{\mc S}^{j_1j_2}$ and $d_{\mc U}^{j_1j_2}$ are similar to those of $d_{TT}^{j_1 j_2}$ and $d_{T\mc O}^{j_1 j_2}$ with minor modification due to the linear superposition of the diagrams.
  \item The computation of the coefficient $d_{TTT}^{j_1 j_2 j_3}$ corresponds to Fig.~\ref{o3}, and the five terms in $d_{TTT}^{j_1 j_2 j_3}$ correspond to the diagrams Fig.~\ref{p5}, \ref{p6}, \ref{p7}, \ref{p8} and \ref{p3}, respectively.
\end{itemize}

In the above discussion, we have been focusing on the quasiprimary operators in the holomorphic part, we can consider the ones in the antiholomorphic part as well. For $CFT_n$, we can count  the number of linear independent quasiprimary operators as
\bea
&& \tr' x^{L_0+\bar L_0}=\lt( 1+n x^2+\f{n(n+1)}{2}x^4+\f{n(n-1)}{2}x^5+\f{n(n+1)(n+5)}{6}x^6+O(x^7) \rt)^2  \nn\\
&& \phantom{\tr' x^{L_0+\bar L_0}}
=1+2nx^2+n(2n+1)x^4+n(n-1)x^5+\f{n(n+1)(4n+5)}{3}x^6+O(x^7).
\eea
So the number of quasiprimary operators at each level $L_0+\bar L_0$ is
\begin{center}
\begin{tabular}{c|c|c|c|c|c|c|c|c}
  $L_0+\bar L_0$ & 0 & 1 & 2 & 3 & 4 & 5 & 6 & $\cdots$  \\\hline
 \# &1 &0 & $2n$ & 0 & $n(2n+1)$ & $n(n-1)$ & $\f{n(n+1)(4n+5)}{3}$ & $\cdots$  \\
\end{tabular}
\end{center}
We do not list these quasiprimary operators explicitly here, as they will not be used  directly in the present work.

\section{Applications}\label{s4}

In this section, using the coefficients we derived above we discuss the R\'enyi entropies in  two examples. The first case is the one short interval on cylinder, and the second case is two intervals on complex plane with small cross ratio.

\subsection{One short interval on cylinder}

In this case the spatial part of the 2D CFT is a circle with length $L$. We choose the coordinate of the cylinder be $z$ and the subsystem $A$ to be an interval $A=[0,\ell]$ with $\ell \ll L$. The R\'enyi entanglement entropy of $A$ is known \cite{Calabrese:2004eu}
\be \label{e50}
S_n=\f{c}{6} \lt( 1+\f{1}{n} \rt) \log \lt( \f{L}{\pi\e}\sin\f{\pi\ell}{L} \rt).
\ee

\begin{figure}
\centering
\subfigure[]{\includegraphics[width=6.2cm]{o0.pdf} \label{hua0}}
\subfigure[]{\includegraphics[width=6.2cm]{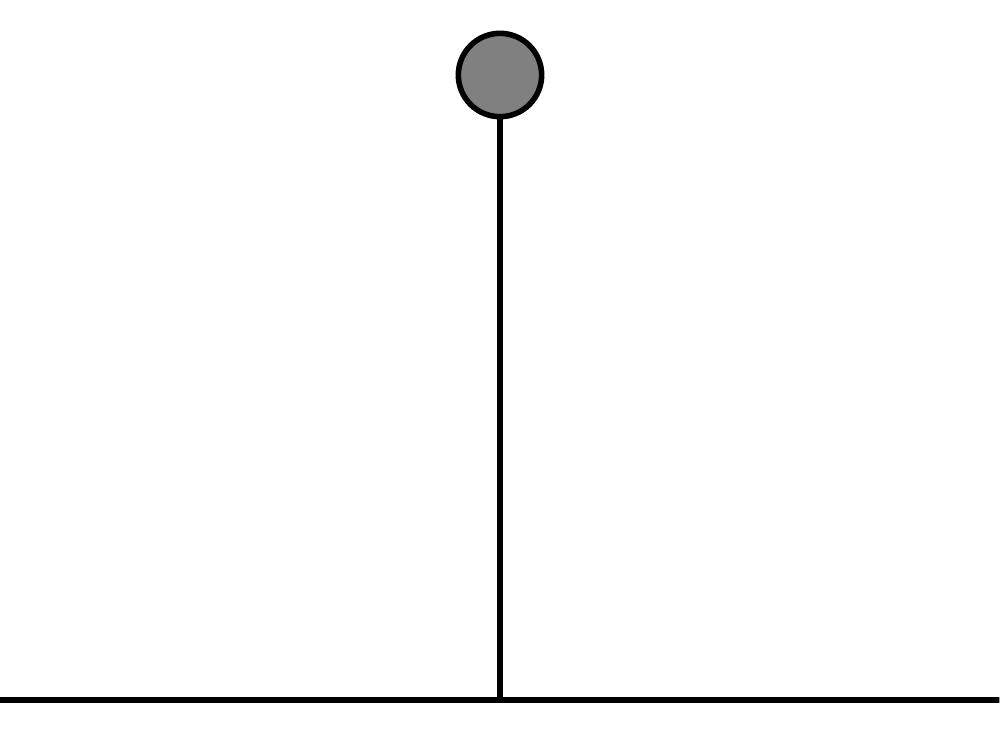} \label{hua1}}
\subfigure[]{\includegraphics[width=6.2cm]{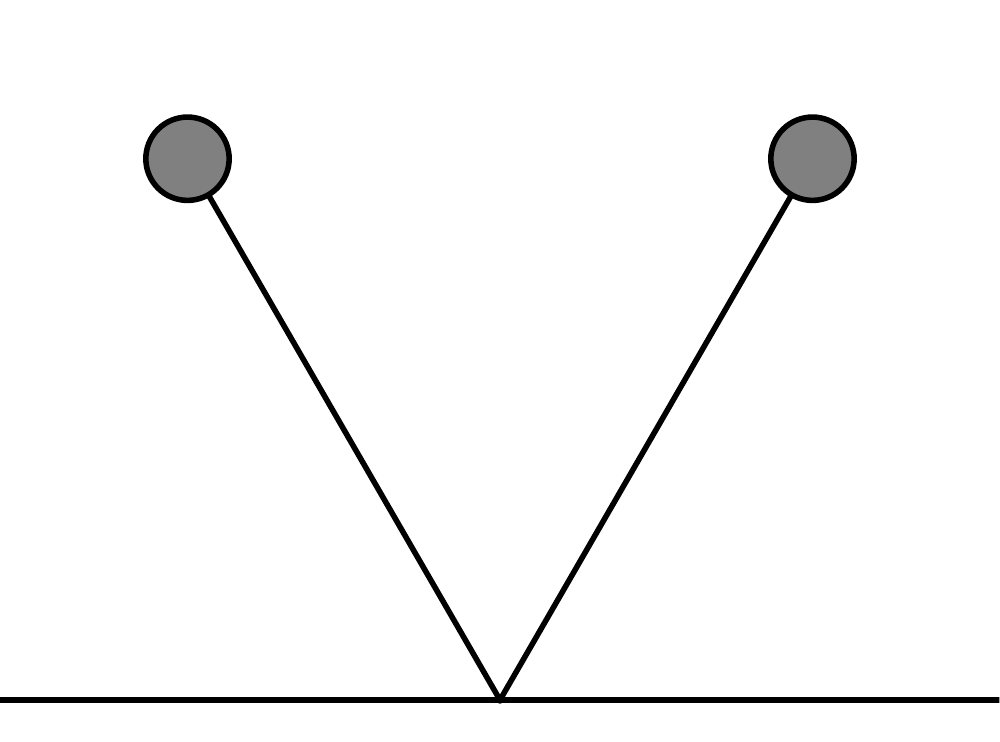} \label{hua2}}
\subfigure[]{\includegraphics[width=6.2cm]{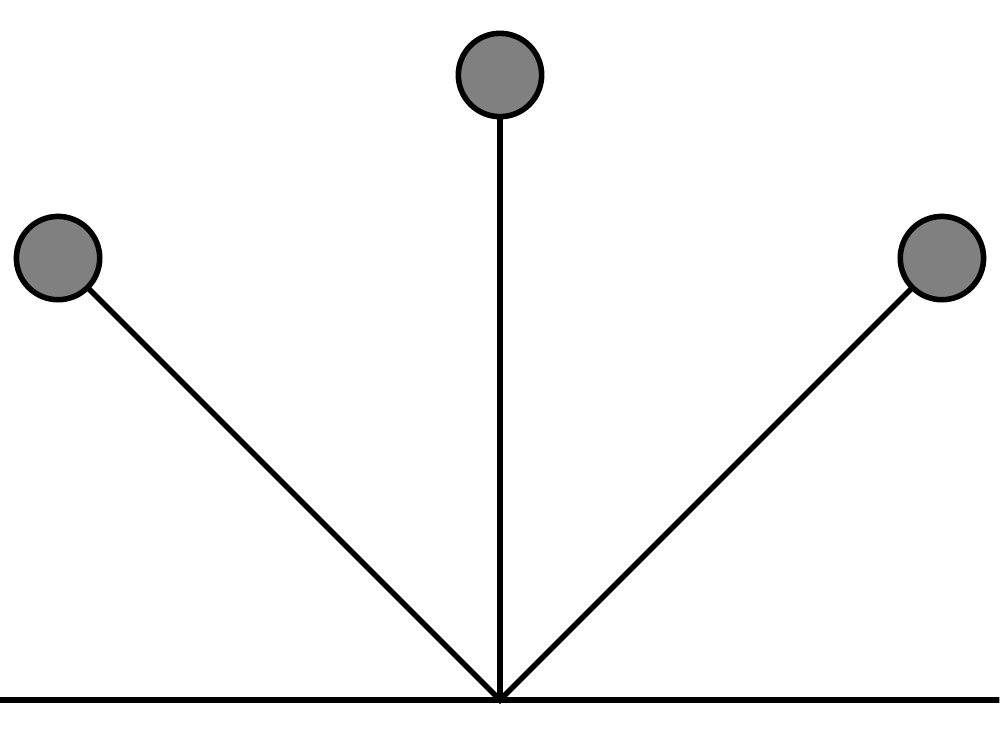} \label{hua3}}
\caption{The OPE of twist operators and the expectation values of the operators on cylinder} \label{hua}
\end{figure}

We can use the replica trick for the CFT on cylinder, and get $CFT_n$. The OPE (\ref{e2}) is a local property of operators and would not be changed by finite length effect. Thus we have
\be \label{j1}
\Tr \r^n_A=\lag \s(\ell,\ell)\td\s(0,0)  \rag_L=c_n \ell^{-\f{c}{6}\lt( n-\f{1}{n} \rt)}\sum_K d_K \ell^{h_K +\bar h_K} \lag \Phi_K(0,0) \rag_L,
\ee
with $K$ being the summation over all the linear independent quasiprimary operators. Note that there is transformation invariance in both directions of the cylinder, so the expectation value of one operator on the cylinder $\lag \Phi_K(z,\bar z) \rag_L$ must be independent of the coordinates, and so the derivative terms vanish uniformly. At odd levels, there is at least one derivative term in one copy of the CFT, and so the expectation value of the operator on cylinder vanish too. Thus we could get the above simple formula ($\ref{j1}$).

As we have mentioned, the OPE of  $T$ and $\bar T$ is trivial, and so the linear independent quasiprimary operators could be written as trivial multiplication of holomorphic and antiholomorphic parts
\be
\Phi_K(z,\bar z)=\Phi_{K_1}(z) \bar\Phi_{\bar K_2}(\bar z).
\ee
In this case we have
\bea \label{e20}
&& \a_K=\a_{K_1}\bar \a_{\bar K_2}, ~~~ \bar \a_{\bar K_2}=\a_{K_2}, \nn\\
&& d_K=d_{K_1}\bar d_{\bar K_2}, ~~~ \bar d_{\bar K_2}=d_{K_2}, \nn\\
&& \lag \Phi_K(z,\bar z) \rag_L=  \lag \Phi_{K_1}(z) \rag_L  \lag\bar\Phi_{\bar K_2}(\bar z)\rag_L, \nn\\
&& \lag\bar\Phi_{\bar K_2}(\bar z)\rag_L=\lag\Phi_{ K_2}( z)\rag_L.
\eea
Then our calculation could be simplified as
\be \label{e34}
\Tr \r^n_A=c_n \ell^{-\f{c}{6}\lt( n-\f{1}{n} \rt)} \lt( \sum_K d_K \ell^{h_K } \lag \Phi_K(0) \rag_L \rt)^2,
\ee
with $K$ being the summation over all the linear independent holomorphic quasiprimary operators.

We use the transformation
\be
z \to f(z)=e^{\f{2\pi i}{L}z}
\ee
which maps the cylinder with the coordinate $z$ to the complex plane with the coordinate $f$. With the transformation we could get
\bea \label{e17}
&& \lag T(z_j) \rag_L=\f{\pi^2c}{6L^2},  \nn\\
&& \lag \mc O(z_j) \rag_L=\f{\pi^4c(5c+22)}{180L^4},  \nn\\
&& \lag \mc Q(z_j) \rag_L=-\f{62\pi^6 c}{945L^6},  \nn\\
&& \lag \mc R(z_j) \rag_L=\f{\pi^6 c(2c-1)(5c+22)(7c+68)}{216(70c+29)L^6}, \nn\\
&& \lag \mc S_{j_1j_2}(z) \rag_L= \lag \mc U_{j_1j_2}(z) \rag_L=0, \nn\\
&& \lag T(z_{j_1})T(z_{j_2})\rag_L=\lt( \f{\pi^2c}{6L^2} \rt)^2,   \nn\\
&& \lag T(z_{j_1})\mc O(z_{j_2})\rag_L=\f{\pi^2c}{6L^2} \f{\pi^4c(5c+22)}{180L^4} ,  \nn\\
&& \lag T(z_{j_1})T(z_{j_2})T(z_{j_3}) \rag_L=\lt( \f{\pi^2c}{6L^2} \rt)^3.
\eea
Then using (\ref{e16}), (\ref{e17}) and (\ref{e34}), we could find the R\'enyi entanglement entropy
\bea \label{j50}
&& S_n=-\f{1}{n-1}\log \Tr \r_A^n  \nn\\
&&\phantom{S_n}
=\f{c}{6} \lt( 1+\f{1}{n} \rt)
\lt( \log\f{\ell}{\e}-\frac{\pi ^2 \ell ^2}{6 L^2}-\frac{\pi ^4 \ell ^4}{180 L^4}-\frac{\pi ^6 \ell ^6}{2835 L^6}
     +O\lt(\f{\ell}{L}\rt)^8 \rt),
\eea
which matches (\ref{e50}) to the order of $O(\ell^6)$.

The computation of the entropy involves the  diagrams Fig.~\ref{hua0}, \ref{hua1}, \ref{hua2} and \ref{hua3}. In (\ref{j1}) we have $C_K=c_n \ell^{-\f{c}{6}\lt( n-\f{1}{n} \rt)+h_K+\bar h_K} d_K$, being the coupling of the OPE vertex operator. The expectation value of an operator on the cylinder $\lag \Phi_K(0,0) \rag_L$ is actually the product of the ones of the operator in each replica of the CFT, with individual expectation value in a replica being represented by a small gray disk in the diagrams. So we actually have only contributions from tree diagrams, and this is in accord with the fact that the final result (\ref{j50}) is proportional to the central charge $c$.

The finite temperature effect is the same to the  finite length case if we substitute $L \to i\b$ with $\b$ being the inverse temperature. We could reproduce the result \cite{Calabrese:2004eu}
\bea
&& S_n=\f{c}{6} \lt( 1+\f{1}{n} \rt) \log \lt( \f{\b}{\pi\e}\sinh\f{\pi\ell}{\b} \rt)  \nn\\
&& \phantom{S_n}
=\f{c}{6} \lt( 1+\f{1}{n} \rt)
\lt( \log\f{\ell}{\e}+\frac{\pi ^2 \ell ^2}{6 \b^2}-\frac{\pi ^4 \ell ^4}{180 \b^4}+\frac{\pi ^6 \ell ^6}{2835 \b^6}
     +O\lt(\f{\ell}{\b}\rt)^8 \rt).
\eea

\subsection{Two intervals on complex plane with small cross ratio}

We consider the case of two short disjoint intervals on the complex plane. We choose $A=[0,y]\cup[1,1+y]$ with $y$ being small, and thus the cross ratio is $x=y^2$. Using (\ref{e2}) and the orthogonality of quasiprimary operators in $CFT_n$ we could get
\bea \label{e49}
&& \Tr \r^n_A=\lag \s(1+y,1+y) \td \s(1,1) \s(y,y)\td \s(0,0)  \rag_C  \nn\\
&& \phantom{\Tr \r^n}=c_n^2(y^2)^{-\f{c}{6}\lt( n-\f{1}{n} \rt)}\sum_K \a_K d_K^2 y^{2(h_K+\bar h_K)} \\
&&\phantom{\Tr \r^n=}
\times \sum_{m,r,p,q\geq0}(-)^{m+r}\f{(m+p)!(r+q)!}{m!r!p!q!}a_K^m \bar a_K^r a_K^p \bar a_K^q
        C_{2h_K+m+p-1}^{m+p} C_{2\bar h_K+r+q-1}^{r+q}y^{m+r+p+q}, \nn
\eea
with $K$ being the summation of all the quasiprimary operators. One can see easily that only even powers of $y$, and so integer powers of $x$, contribute. Also using the arguments around (\ref{e20}), the above formula could be simplified as
\bea
&& \Tr \r^n_A=c_n^2y^{-\f{c}{3}\lt( n-\f{1}{n} \rt)} \lt( \sum_K \a_K d_K^2 y^{2h_K}
       \sum_{m,p\geq0}(-)^{m}\f{(m+p)!}{m!p!}a_K^m a_K^p
        C_{2h_K+m+p-1}^{m+p} y^{m+p} \rt)^2,
\eea
with $K$ being the summation over all linear independent holomorphic quasiprimary operators.

After some quite nontrivial summations\footnote{Some useful formulas used in the summation have been gathered into Appendix~\ref{sa}.} we get the mutual information
\bea \label{e38}
&&I_n=\f{c}{3}(1+\f{1}{n})\log\f{y}{\e}+\f{1}{n-1}\log\Tr\r_A^n, \nn\\
&&\phantom{I_n}
     =I_n^{tree}+I_n^{1-loop}+I_n^{2-loop}+\cdots.
\eea
The tree part, or the so-called classical part, being proportional to the central charge $c$, is
\bea
&& I_n^{tree}=\frac{c (n-1) (n+1)^2 x^2}{144 n^3}+\frac{c (n-1) (n+1)^2 x^3}{144 n^3}+\frac{c (n-1) (n+1)^2 \left(1309 n^4-2 n^2-11\right) x^4}{207360 n^7}  \nn\\
&& \phantom{I_n^{cl}=}
+\frac{c (n-1) (n+1)^2 \left(589 n^4-2 n^2-11\right) x^5}{103680 n^7}\nn\\
&& \phantom{I_n^{cl}=}
+\frac{c (n-1) (n+1)^2 \left(805139 n^8-4244 n^6-23397 n^4-86 n^2+188\right) x^6}{156764160 n^{11}}+O\left(x^7\right)
\eea
This matches the result in \cite{Headrick:2010zt,Hartman:2013mia,Faulkner:2013yia}. The quantum 1-1oop part, being proportional to $c^0$, is
\bea
&& I_n^{1-loop}=\frac{(n+1) \left(n^2+11\right) \left(3 n^4+10 n^2+227\right) x^4}{3628800 n^7} \nn\\
&& \phantom{I_n^{1-loop}=}
+\frac{(n+1) \left(109 n^8+1495 n^6+11307 n^4+81905 n^2-8416\right) x^5}{59875200 n^9}\nn\\
&& \phantom{I_n^{1-loop}=}
+\frac{(n+1) \left(1444050 n^{10}+19112974 n^8+140565305 n^6+1000527837 n^4-167731255 n^2-14142911\right) x^6}{523069747200 n^{11}} \nn\\
&& \phantom{I_n^{1-loop}=}+O\left(x^7\right),
\eea
and this matches the result in\cite{Headrick:2010zt,Barrella:2013wja}. Remarkably there is also the quantum 2-loop contribution, being proportional to $1/c$,
\be
I_n^{2-loop}=\frac{(n+1) \lt(n^2-4\rt) \left(19 n^8+875 n^6+22317 n^4+505625 n^2+5691964\right) x^{6}}{70053984000 n^{11} c}+O\left(x^{7}\right),
\ee
and this is a new result. Here we have classified the contributions according to the order of the inverse of central charge $\f{1}{c}$, which in the large $c$ limit corresponds to tree, 1-loop, and 2-loop contributions  in the gravity side\cite{Headrick:2010zt}.

\begin{figure}
\centering
\subfigure[trivial]{\includegraphics[width=6cm]{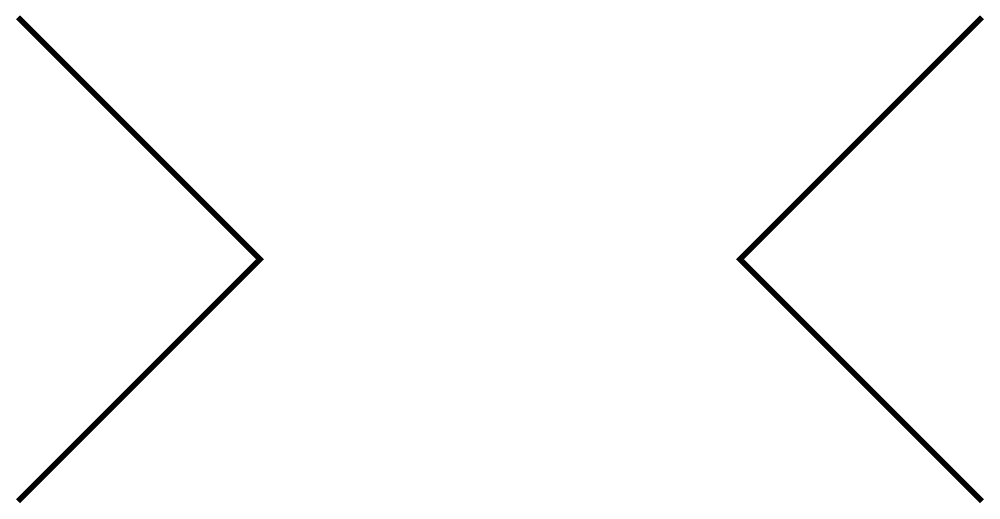} \label{l-1}}
\subfigure[tree]{\includegraphics[width=6cm]{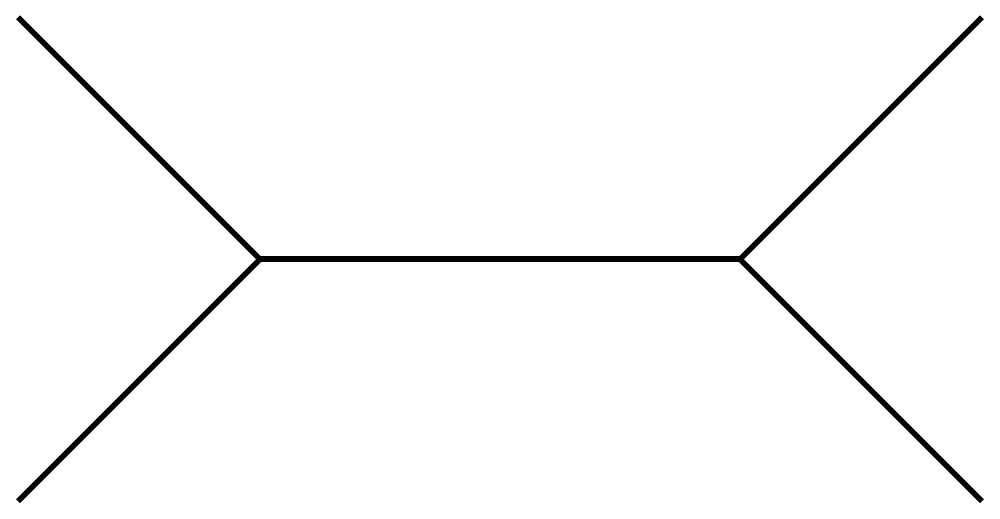} \label{l0}}
\subfigure[1-loop]{\includegraphics[width=6cm]{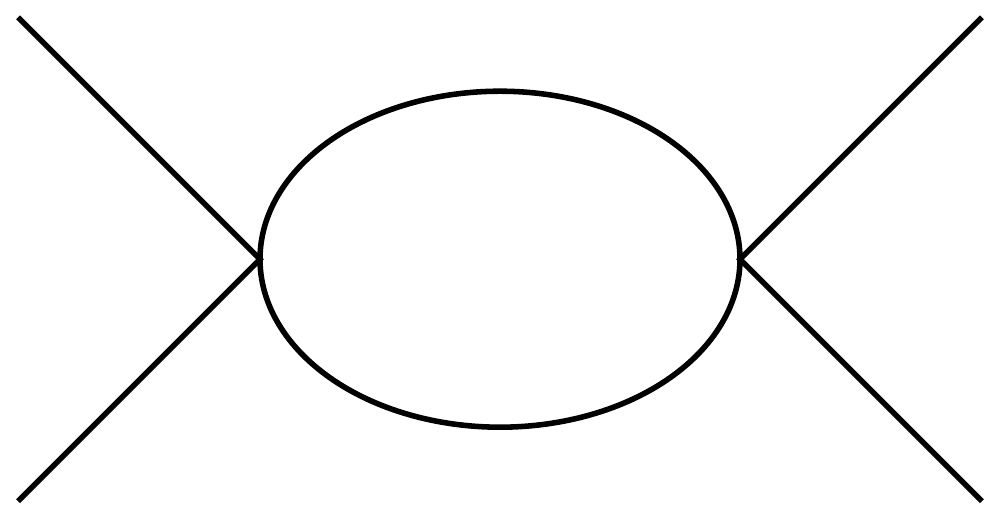} \label{l1}}
\subfigure[2-loop]{\includegraphics[width=6cm]{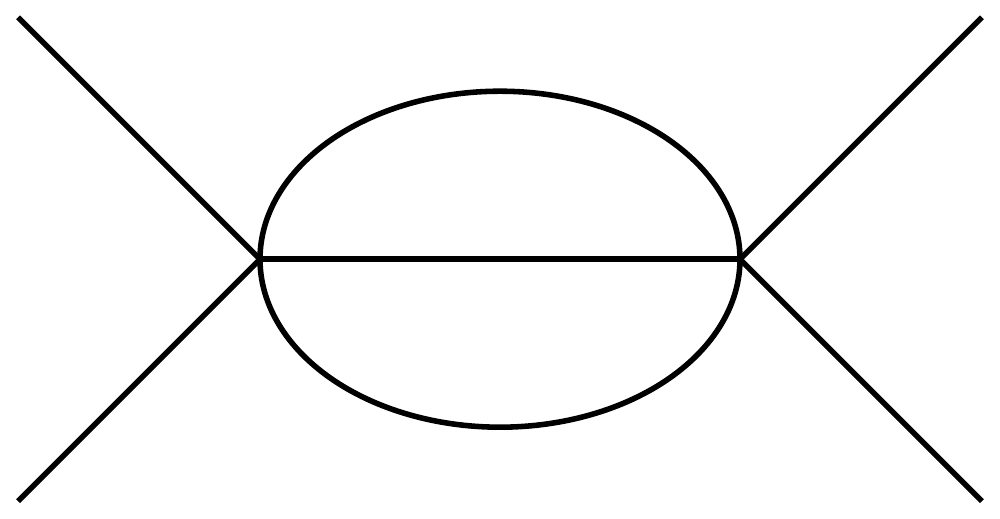} \label{l2}}
\caption{The diagrams for two intervals} \label{f5}
\end{figure}

The computations in (\ref{e49}) to order $x^6$ corresponds to the summation of the diagrams Fig.~\ref{l-1}, \ref{l0}, \ref{l1} and \ref{l2}. In the figures Fig.~\ref{f5}, the four external lines represents the four twist operators inserted at the boundary of the two intervals. The two vertexes represent the OPE coefficients of the twist operators at each interval, and the internal lines that connect the two vertexes could be understood as the propagators. Each term in the summation of (\ref{e49}) could be represented by a diagram like this. For example a typical term involves the OPE coefficients of the operators $\p^m\bar \p^r \Phi_K$ and $\p^p\bar \p^q \Phi_K$
\bea
&& C_K^{(m,r)}=c_n y^{-\f{c}{6}\lt( n-\f{1}{n} \rt)+m+r}  d_K \f{a_K^m}{m!}\f{\bar a_K^r}{r!},  \nn\\
&& C_K^{(p,q)}=c_n y^{-\f{c}{6}\lt( n-\f{1}{n} \rt)+p+q}  d_K \f{a_K^p}{p!}\f{\bar a_K^q}{q!},
\eea
as well as the propagator
\be
\a_K (-)^{m+r} (m+p)!(r+q)! C_{2h_K+m+p-1}^{m+p} C_{2\bar h_K+r+q-1}^{r+q}.
\ee

Note that we are a little sloppy in naming the figures Fig.~\ref{f5}. We have called the diagrams Fig.~\ref{l0}, \ref{l1} and \ref{l2} as the tree, 1-loop, and 2-loop diagrams respectively by their appearances, but they are not in one to one correspondence to the tree, 1-loop, and 2-loop contributions of the mutual information in (\ref{e38}). In the above computations we have two quantities to be expanded, and explicitly we expand firstly in terms of small cross ratio $x$ and then in terms of small inverse of the central charge  $\f{1}{c}$. In the expansion of $x$, it is easy to see that the contribution of the $l$-loop diagrams starts at order $x^{2(l+1)}$. Thus if we want the mutual information (\ref{e38}) to the order $x^{2m}$ or $x^{2m+1}$, we need only consider the diagrams with number of loops $l \leq m-1 $. However, in the expansion of $\f{1}{c}$ the contributions of the each diagram is not clear. We cannot consider these diagrams separately, and to get the general correct result we have to sum the contributions of all the diagrams to some order of $x$. It seems to be true that the most leading contribution to the $\f{1}{c^p}$ terms of the mutual information starts from diagrams with number of loops $l=p+1$ and is of the order $x^{2(p+2)}$. There may be some other nontrivial relations between the order of $\f{1}{c}$ and the number of loops of the diagrams, but we cannot fix this problem with the limited orders of $x$ we have calculated.


The agreement of the tree level and 1-loop contribution with the known result in the gravity is remarkable, especially considering the complexity of the computation. Firstly note that the dependence of the coefficients $\a_K$ and $d_K$ on the central charge $c$ is not linear, but the final results could be organized neatly according to the order of $c$. Actually   there are many nontrivial cancelations in the calculation that give a relatively simple final result. We wish that there could be other more effective ways of the computation.



\section{Conclusion and discussion}\label{s5}

The recent study of the R\'enyi entropy and its 1-loop quantum correction in the AdS$_3$ gravity sheds new light on the AdS$_3$/CFT$_2$ correspondence. In this case, the quantitative comparison between two sides is feasible. In the CFT side, even for the multiple disjoint intervals, the entropies could be computed in the short interval limit, using the OPE of twist operators.
In this paper we developed the short interval expansion of twist operators by considering the derivatives of the quasiprimary operators. This allowed us to get the subleading contributions of R\'enyi entropy. We only considered the  contributions of the operators constructed using the stress tensors. We calculated the  expansion coefficients of the quasiprimary operators to level 6. These coefficients could be used to compute  several leading contributions of R\'enyi entropy of one interval on cylinder and two intervals on complex plane. In the large central charge limt, the results are in perfect matches with the known results in the bulk. For a short interval on cylinder, we could get the finite size correction to the order $\ell^6$, and for two intervals on complex plane we could reproduce the classical, 1-loop results to order $x^6$. Moreover to order $x^6$  we found the contributions of order $1/c$, which correspond to the 2-loop corrections in gravity.

One import lesson from our investigation is that the R\'enyi entropy opens a new window to study the AdS$_3$/CFT$_2$ correspondence. In the case of two disjoint intervals, the R\'enyi entropy $S_2$ is just the partition function on a torus with a modular parameter. This partition function corresponds to the
1-loop determinant of physical fluctuations around the thermal AdS space. In this case, the partition function encodes the information of the spectrum and has been used as the check of the correspondence. The higher R\'enyi entropy $S_n, n>2$ present new challenges and criterion. For example, it has been conjectured for a long time that the pure AdS$_3$ gravity could be holomorphically dual to a 2D CFT, possibly a Liouville field theory\cite{Brown:1986nw,Witten:1988hc,Carlip:2005zn,Maloney:2007ud}. We have shown that to order 6, the contributions from vacuum Verma module are in perfect match with the pure gravity results. This puts strong constraint on the  CFT dual. At first looking, the Liouville field theory seems have much richer spectrum than required. However, these spectrum may correspond to nonperturbative objects in the bulk.

Our investigations in this work could be extended in several directions.
\begin{itemize}
\item First of all, it would be interesting to compute the R\'enyi entropy of a concrete CFT model, considering the limited knowledge on this issue.\footnote{Note that the leading order contributions to the R\'enyi entropy in the minimal models have been computed in \cite{Rajabpour:2011pt}.} In general, we have to include the contributions from other conformal families, besides the ones from vacuum Verma module which is the conformal family of the identity operator. In this case the two intervals on complex plane result (\ref{e38}) will be changed, but the one interval on cylinder case (\ref{j50}) is exact and should not change. This is expected, since a primary operator that is not the identity has vanishing expectation value on a cylinder.
\item Secondly, it would be interesting to study the AdS$_3$/CFT$_2$ correspondence with other matter coupling. In particular, the R\'enyi entropy may provide another window to check the minimal model holography in \cite{Gaberdiel:2012uj}.
\item Thirdly, it would be worthwhile to discuss the R\'enyi entropy in the gravity with higher derivative corrections \cite{deBoer:2011wk,Hung:2011xb,Bhattacharyya:2013jma,Chen:2013qma,Fursaev:2013fta,Bhattacharyya:2013gra}.
\item It would be nice to generalize our study to the case with more than two intervals. This is workable but tedious.
\item It is certainly important to generalize our prescriptions to higher dimensions. We hope that those generalizations will be done in the future.
\end{itemize}

\vspace*{1cm}
\noindent {\large{\bf Acknowledgments}}\\
We would like to thank Xi Dong and Sean Hartnoll for valuable correspondence, and we also thank Jiang Long and Jie-qiang Wu for helpful discussions. Special thanks to Matthew Headrick for his powerful Mathematica code Virasoro.nb that could be downloaded at his personal homepage \url{http://people.brandeis.edu/~headrick/Mathematica/index.html}. The work was in part supported by NSFC Grant No. 11275010, No. 11335012 and No. 11325522. JJZ was also in part supported by the Scholarship Award for Excellent Doctoral Student granted by the Ministry of Education of China.
\vspace*{1cm}

\begin{appendix}

\section{Some useful formulas} \label{sa}

In the appendix we summarize some formulas that is needed in our calculation. We define
\be
f_m(n)=\sum_{j=1}^{n-1}\f{1}{ \lt( \sin\f{\pi j}{n} \rt)^{2m}},
\ee
and explicitly we need
\bea
&& f_1(n)=\frac{n^2-1}{3}, ~~~ f_2(n)=\frac{(n^2-1) \left(n^2+11\right)}{45} , ~~~
f_3(n)=\frac{(n^2-1)  \left(2 n^4+23 n^2+191\right)}{945} ,  \nn\\
&& f_4(n)=\frac{(n^2-1) \left(n^2+11\right) \left(3 n^4+10 n^2+227\right)}{14175},  \nn\\
&& f_5(n)=\frac{(n^2-1) \left(2 n^8+35 n^6+321 n^4+2125 n^2+14797\right)}{93555}, \\
&& f_6(n)=\frac{(n^2-1) \left(1382 n^{10}+28682 n^8+307961 n^6+2295661 n^4+13803157 n^2+92427157\right)}{638512875}. \nn
\eea
The above formulas are useful because it often appears in the calculation that
\be
\sum_{0\leq j_1 <j_2 \leq n-1}\f{1}{s^{2m}_{j_1j_2}}=\f{n}{2}f_m(n).
\ee
There are also several summation formulas listed below.
\bea
&& \sum_{0\leq j_1 <j_2<j_3 \leq n-1} \f{1}{s^2_{j_1j_2}s^2_{j_2j_3}s^2_{j_1j_3}}=
   \frac{n\lt(n^2-1\rt)\lt(n^2-4\rt) \left(n^2+47\right)}{2835},  \nn\\
&& \sum_{0\leq j_1 <j_2<j_3 \leq n-1} \f{1}{s^4_{j_1j_2}s^4_{j_2j_3}s^4_{j_1j_3}}=
   \frac{n\lt(n^2-1\rt)\lt(n^2-4\rt) \left(19 n^8+875 n^6+22317 n^4+505625 n^2+5691964\right)}{273648375},  \nn\\
&& \sum_{0\leq j_1 <j_2<j_3 \leq n-1} \lt( \f{1}{s^4_{j_1j_2}}+\f{1}{s^4_{j_2j_3}}+\f{1}{s^4_{j_1j_3}} \rt)=
   \frac{n(n^2-1)(n-2) \left(n^2+11\right)}{90},  \\
&& \sum_{0\leq j_1 <j_2<j_3 \leq n-1} \lt( \f{1}{s^4_{j_1j_2}}+\f{1}{s^4_{j_2j_3}}+\f{1}{s^4_{j_1j_3}} \rt)^2=
   \frac{n  (n^2-1) (n-2) \left(n^2+11\right) \left(3 n^4+8 n^3+26 n^2+152 n+531\right)}{28350}, \nn\\
&& \sum_{0\leq j_1 <j_2<j_3 \leq n-1} \f{1}{s^2_{j_1j_2}s^2_{j_2j_3}s^2_{j_1j_3}}
   \lt( \f{1}{s^4_{j_1j_2}}+\f{1}{s^4_{j_2j_3}}+\f{1}{s^4_{j_1j_3}} \rt)=
   \frac{n (n^2-1) (n^2-4) \left(3 n^6+125 n^4+1757 n^2+21155\right)}{467775}. \nn
\eea

\end{appendix}



\providecommand{\href}[2]{#2}\begingroup\raggedright\endgroup

\end{document}